\pdfoutput=1
\RequirePackage{ifpdf}
\documentclass[paper,notoc]{JHEP3}
\usepackage{amssymb}   % for math
\usepackage{amsmath}
\usepackage{multirow}
\usepackage{graphicx}
\usepackage{booktabs} % for book-like tables
\usepackage{cite}
 \usepackage{url}\usepackage{epstopdf}
\allowdisplaybreaks
\newcommand{\ord}{\begin{cal}O\end{cal}}

\def\beq{\begin{equation}}
\def\eeq{\end{equation}}
\def\bea{\begin{eqnarray}}
\def\eea{\end{eqnarray}}
\def\bsp#1\esp{\begin{split}#1\end{split}}

\def\beq{\begin{equation}}
\def\eeq{\end{equation}}
\def\bsp#1\esp{\begin{split}#1\end{split}}

\newcommand{\nnnlo}[0]{N$^3$LO }

\title{
CP-even scalar boson production via gluon fusion at the LHC
}

\author{Charalampos Anastasiou$^a$, Claude Duhr$^{b,c}$\footnote{On leave from the `Fonds National de la Recherche Scientifique' (FNRS), Belgium.}, Falko Dulat$^a$,
  Elisabetta Furlan$^{a,d}$, Thomas Gehrmann$^{e}$,
Franz Herzog$^f$,
Achilleas Lazopoulos$^a$,
Bernhard Mistlberger$^b$\\
{}$^a$Institute for Theoretical Physics, ETH Z\"urich,
  8093 Z\"urich, Switzerland
\\
{}$^b$Theoretical Physics Department, CERN, Geneva, Switzerland\\
{}$^c$Center for Cosmology, Particle Physics and Phenomenology (CP3),\\
\phantom{{}$^c$}Universit\'e catholique de Louvain,\\
\phantom{{}$^c$}Chemin du Cyclotron 2, 1348 Louvain-La-Neuve, Belgium\\
{}$^d$Kavli Institute for Theoretical Physics, University of California, Santa Barbara, CA 93106, USA
\\
{}$^e$Physik-Institut, Universit\"at Z\"urich,
Winterthurerstrasse 190, 8057 Z\"urich, Switzerland
{}$^f$Nikhef, Science Park 105, NL-1098 XG Amsterdam, The Netherlands
}

\preprint{CP3-16-12, ZU-TH-11/16, NIKHEF 2016-013, CERN-TH-2016-085, NSF-KITP-16-039}

\abstract{
In view of the searches at the LHC for scalar particle resonances
in addition to the 125 GeV Higgs boson, we present the cross section
for a CP-even scalar produced via gluon fusion at \nnnlo in
perturbative QCD assuming that it couples directly to gluons
in an effective theory approach. We refine
our prediction by taking into account the possibility that the scalar
couples to the top-quark and computing the corresponding contributions
through NLO in perturbative QCD.
We assess the theoretical uncertainties of the cross section
due to missing higher-order QCD effects and we provide the necessary
information for obtaining the cross section value and uncertainty from
our results in specific scenarios beyond the Standard Model. We also give detailed results
for the case of a 750~GeV scalar, which will be the subject of intense experimental studies.
}

\keywords{Higgs physics, QCD, gluon fusion, BSM, scalar boson}

\begin{document}
\section{Introduction}
\label{sec:intro}

The Higgs boson is the only elementary  scalar particle of the Standard Model (SM) and the
first such particle whose existence has been confirmed experimentally~\cite{Aad:2012tfa,Chatrchyan:2012xdj}.
Many well-motivated extensions of the SM predict additional electrically neutral and
colourless scalar particles (elementary or not). Experimental searches to hunt for
these hypothetical new particles are thus of extreme importance.
The LHC experiments have the potential to discover new scalar resonances by using similar techniques
and signatures as in the search for the SM Higgs boson. In particular, in this context both
the ATLAS an CMS experiments have recently reported an excess in the diphoton mass spectrum
at an invariant mass of $\sim 750$~GeV~\cite{ATLAS_gaga, CMS:2015dxe}.
Even if it is still premature to speculate if the source of this excess is indeed a new resonance or
merely a statistical fluctuation, this excess is exemplary for the kind of signals that may arise from
the production of heavy singlet scalars in the current run of the LHC.

While specific models with colour-neutral scalars may differ drastically in the principles which
motivate them as well as in their spectra and interactions, the computation of
QCD radiative corrections to their production cross section could share many common features
that are independent of the underlying model.
In particular, the production of hypothetical Higgs-like scalars at
the LHC may be favourable in the gluon-fusion process due to the large
value of the gluon-gluon luminosity.  In this case, the model-dependence of the
production cross section may only enter through the coupling describing the (effective)
interaction of the Higgs-like scalar with the gluons and quarks.

The purpose of this paper is to provide benchmark cross sections
for a colorless CP-even scalar in an effective theory which is similar to the one
obtained for the Higgs boson in the Standard Model after
integrating out the top-quark. The cross section for the scalar can
be obtained from the cross section of the Higgs boson for a wide range of models, by
appropriately taking into account the corresponding Wilson coefficient.
Our computation is valid through \nnnlo in perturbative QCD and
is based on recent results for the SM Higgs boson gluon-fusion cross section to that
order~\cite{Baikov:2009bg,Gehrmann:2010ue,Hoschele:2012xc, Anastasiou:2012kq, Anastasiou:2013srw,Anastasiou:2013mca, Li:2013lsa, Anastasiou:2014vaa, Anastasiou:2014lda, Dulat:2014mda, Duhr:2014nda, Li:2014afw,  Anastasiou:2015ema, Anzai:2015wma, Anastasiou:2016cez, Anastasiou:2015yha}.
We consider mass values for the new scalar particle ranging from 10~GeV to 3~TeV,
which is the same range as the one targeted by the searches of ATLAS and CMS
in the Run 2 of the LHC~\cite{HXSWG}.

In some extensions of the Standard Model (see, for example, ref.~\cite{Dolan:2016eki}
and references therein) the new Higgs-like scalar has a non-negligible coupling to the
Standard Model top-quark.
For scalar masses below the top-pair threshold the Wilson coefficient in our effective
field theory (EFT) can
be made to account for a coupling of the scalar to the top-quark. However, for
scalar masses around and above threshold, the effective theory
approach is not accurate.  For this purpose, we extend the effective
theory by adding a Yukawa-type interaction of the top-quark with the
scalar and compute the additional contributions generated by this coupling.

While we can correct our \nnnlo cross-section in the effective theory for
contributions of light Standard Model particles through NLO in QCD,
it is not possible to do so in a model-independent way for contributions
from new relatively light particles with a mass smaller than about half the mass
of the scalar. The presence of such particles will invalidate our effective-theory
computation. To estimate this effect, we consider the difference between the
effective theory NLO cross-section
and the exact NLO cross-section for a 750~GeV scalar which couples
to a new top-like quark, as a function of the mass of the quark.
We find that the two cross-sections can differ
by more than the QCD scale uncertainty for top-like quark masses as heavy as twice the mass of the scalar.
Nevertheless, we observe that one can still make use of the effective theory computation
for such a model also for lower values of the quark mass in order to estimate the relative
size of the QCD corrections ($K-$~factor).

This article is organised as follows. In section~\ref{Sec:N3_EFT}
we present the effective theory setup of our calculation
and discuss how the results can be adapted to suite a large class of models through
the Wilson coefficient. We also discuss
the theoretical uncertainty that is associated with the N$^3$LO QCD corrections in the
effective theory. In section~\ref{sec:width},
we discuss the inclusion of finite width effects in the calculation and provide a fit of
the zero-width cross section to enable
the adaptation of our results to models where the scalar resonance is expected
to have a non-vanishing but narrow width.
In section~\ref{Sec:EFT_vs_exact}, we investigate the validity of our effective theory approach
under the assumption that
the coupling between the scalar and the gluons is mediated by a heavy top partner of varying mass.
In section~\ref{sec:finite-top} we go beyond a purely effective description of the scalar
interaction and allow for a non-vanishing coupling
between the Standard Model top and the scalar, for which we present the NLO QCD corrections.
We conclude in section~\ref{sec:concl} and provide appendices with tables of cross section
values for a wide range of masses of the scalar.

\section{\nnnlo QCD corrections in the effective theory approach}
\label{Sec:N3_EFT}

Let us consider a model where the SM is extended by a colourless singlet scalar $S$
of mass $m_S$ and width $\Gamma_S$,
which only couples to the SM in a minimal way. For now, we assume that the only SM
fields that the new scalar
$S$ couples to are the gluons, through an effective operator of dimension five. We will discuss Yukawa couplings to heavy quarks in Section~\ref{sec:finite-top}.
Such a model can be described by a Lagrangian of the form
\beq\label{eq:L_eff}
\mathcal{L}_{\text{eff}}=\mathcal{L}_{\textrm{SM}}+\mathcal{L}_S-\frac{1}{4 v}C_{S} \, S\, G_{\mu\nu}^a G_a^{\mu\nu},
\eeq
where $\mathcal{L}_S$ collects the kinetic term and the potential of the scalar $S$. The vacuum expectation value $v$ is~\cite{PDG},
$$v = \frac{1}{\sqrt{\sqrt{2} G_f}} = 246.22~\textrm{GeV},$$
and $C_S$ is a Wilson coefficient parametrizing the strength of the
coupling of the particle $S$ with the gluons. Except for the value of the Wilson
coefficient, this is the same dimension-five operator as the one coupling the SM Higgs boson
to the gluons, after the top-quark has been integrated out.

Since the scalar $S$ couples to gluons, it can be produced at hadron colliders.
Its hadronic production cross section can be cast in the form
\begin{equation}
\sigma_S(m_S,\Gamma_S,\Lambda_{\rm UV})
= \left| C_S(\mu, \Lambda_{\rm UV}) \right|^2 \eta(\mu,m_S,\Gamma_S)\,,
\label{eq:factorization}
\end{equation}
where $\mu$  is the mass scale introduced by dimensional
regularisation and $\Lambda_{\rm UV}$ is the scale of new physics,
representing collectively the masses of the heavy particles
in some ultraviolet (UV) completion of the theory.
At all orders in perturbation theory, the cross section is independent
of the arbitrary scale $\mu$, with the scale variation of the square of the
Wilson coefficient cancelling the one of the matrix-elements
$\eta$. Since $\eta$ does not depend on $\Lambda_{\rm UV}$,
the scale dependence of the Wilson coefficient itself
is universal, and it does not depend on the (renormalizable) UV completion of the
effective theory.  In particular, the renormalization group evolution equation that
relates the Wilson coefficient at two different scales reads \cite{Spiridonov:1988md}:
\begin{equation}
\frac{C_S(\mu,\Lambda_{\rm UV})}{C_S(\mu_0,\Lambda_{\rm UV})} = \frac{\beta(\mu)}{\beta(\mu_0)}\frac{\alpha(\mu_0)}{\alpha(\mu)}\,,
\end{equation}
with $\beta(\mu)$ the QCD $\beta$-function. We see that, as expected, the evolution equation
only depends on the low-energy effective theory, and it is independent of the details
of the UV completion. As a consequence, if we divide the production cross section by
the value of the Wilson coefficient at some reference scale $\mu_0$,
\begin{equation}
\label{eq:sigma_over_wilson}
\frac{\sigma_S(m_S,\Gamma_S,\Lambda_{\rm UV})}{\left| C_S(\mu_0, \Lambda_{\rm UV})\right|^2} =
\left[\frac{\beta(\mu)}{\beta(\mu_0)} \frac{\alpha(\mu_0)}{\alpha(\mu)}\right]^2 \eta(\mu,m_S,\Gamma_S)\,,
\end{equation}
we obtain a universal ratio that does not depend on the UV completion
(up to corrections due to differences that the various high-energy theories
may induce on the width of the scalar
particle $S$). In particular, since the dimension-five operator in eq.~\eqref{eq:L_eff}
is the same as the dimension-five operator mediating Higgs production in gluon fusion in the SM,
we can
%apply the same reasoning to gluon fusion. This allows us to
extract the right-hand side of
eq.~\eqref{eq:sigma_over_wilson} from Higgs production in gluon-fusion,
%. We can then write
\begin{equation}
\label{eq:sigmaS_over_sigmaH}
\sigma_S(m_S,\Gamma_S,\Lambda_{\rm UV}) =
\left|\frac{ C_S(\mu_0, \Lambda_{\rm UV})}{ C(\mu_0, m_{t})}\right|^2\,
\sigma_H(m_S,\Gamma_S,m_t) \; .
\end{equation}
Here $\sigma_H$ is the gluon-fusion production cross section of a Higgs boson of
mass $m_S$ and width $\Gamma_S$,
in a variant of the SM with $N_f=5$ massless
quarks and the top-quark integrated out, $m_t$ is the top-quark mass and $C(\mu_0,m_{t})$
is the Wilson coefficient multiplying the SM gluon-fusion operator
evaluated at our reference scale $\mu_0$. In the $\overline{\rm MS}$
scheme it reads~\cite{Chetyrkin:1997un,Schroder:2005hy}
\begin{eqnarray}
C(\mu, m_t) & = & - \frac{a_s}{3} \Bigg\{ 1 +a_s\,\frac{11}{4}
+ a_s^2 \left[\frac{2777}{288} - \frac{19}{16} \log\left(\frac{m_t^2}{\mu^2}\right) -
 N_f\left(\frac{67}{96}+\frac{1}{3}\log\left(\frac{m_t^2}{\mu^2}\right)\right)
\right]
 \nonumber \\
  &&+ a_s^3\Bigg[-\left(\frac{6865}{31104}
  + \frac{77}{1728} \log\left(\frac{m_t^2}{\mu^2}\right)
  + \frac{1}{18}\log^2\left(\frac{m_t^2}{\mu^2}\right)\right) N_f^2 \\
  && + \left(\frac{23}{32} \log^2\left(\frac{m_t^2}{\mu^2}\right) -
  \frac{55}{54} \log\left(\frac{m_t^2}{\mu^2}\right)+\frac{40291}{20736}
 - \frac{110779}{13824} \zeta_3 \right) N_f \nonumber\\
  &&  -\frac{2892659}{41472}+\frac{897943}{9216}\zeta_3
  + \frac{209}{64} \log^2\left(\frac{m_t^2}{\mu^2}\right)
  - \frac{1733}{288}\log\left(\frac{m_t^2}{\mu^2}\right)\Bigg] +\ord(a_s^4) \Bigg\} \, ,\nonumber
\end{eqnarray}
with $a_s = \alpha_s(\mu)/\pi$, $\alpha_s$ being the strong coupling
constant in QCD with $N_f=5$ flavors.
We note at this point that eq.~\eqref{eq:sigmaS_over_sigmaH} is valid to all orders in
perturbation theory, and so neither $\sigma_S$ nor $\sigma_H$ depend on the arbitrary scales $\mu_0$ and $\mu$.
They do, however, acquire a dependence on the scale $\mu$ once the perturbative
series is truncated at some finite order. It is of course possible to further separate $\mu$  into
a factorization and a renormalization scale, $\mu_f$ and $\mu_r$, as it is customary,  and
we will assume in the following that this is done implicitly. In addition, $\sigma_S$ also acquires
a dependence on our reference scale $\mu_0$ order by order in perturbation theory, resulting
from the truncation of the perturbative expansion of the ratio of Wilson coefficients.

Equation~\eqref{eq:sigmaS_over_sigmaH} is the basic formula which
allows us to predict the hadronic production cross section of a generic CP-even
singlet scalar through gluon-fusion to high order in perturbative QCD. Indeed, the
gluon-fusion cross section is known through N$^3$LO in QCD in the large-$m_t$
limit~\cite{Anastasiou:2016cez} (for $\Gamma_S=0$) as an expansion around the
Higgs threshold,
and eq.~\eqref{eq:sigmaS_over_sigmaH} easily allows us to convert the result for the Higgs boson
into the corresponding N$^3$LO cross section for a generic CP-even scalar.

At this point we need to make a small technical comment:
both the ratio of Wilson coefficient and the Higgs production cross section on the right-hand side of
eq.~\eqref{eq:sigmaS_over_sigmaH} admit a perturbative expansion up to N$^3$LO.
Multiplying these two perturbative expansions, as suggested by eq.~\eqref{eq:sigmaS_over_sigmaH},
introduces additional terms into the perturbative expansion which are of N$^4$LO accuracy or higher,
thus beyond the reach and the control of our N$^3$LO computation.
In ref.~\cite{Anastasiou:2016cez} it was shown that the numerical impact
of these terms is captured by the scale variation at N$^3$LO if the central scale is
chosen as $\mu=m_S/2$,
and the effects of such terms are therefore accounted for by the scale variation uncertainty assigned to our prediction.

A second comment regards the dependence of the Standard Model Wilson
coefficient on the various scales.
The SM Wilson coefficient $C(\mu_0, m_{t})$ depends on the reference
scale $\mu_0$ and
the top-quark mass $m_t$.
While the value of $\mu_0$ is arbitrary and can be chosen freely,
in practise we choose the value of $\mu_0$ to be the mass of the scalar,
$\mu_0=m_S$, and we set the value of the mass of the top quark to its physical value.
In doing so, we may introduce large logarithms into the computation, order by order in perturbation theory.
For example, we see that at NNLO and \nnnlo the
Wilson coefficient contains a logarithm of the form $\log(m_t/\mu_0)$. Similarly,
the Wilson coefficient expression for the scalar $C_S(\mu_0,
\Lambda_{\rm UV})$  will contain logarithms of the  type
$\log(\Lambda_{\rm UV}/\mu_0)$. In order to
avoid large logarithms due to widely disparate scales in the ratio of
the two Wilson coefficients, it may be preferable to evaluate the SM
Wilson coefficient with a top-quark mass value
$m_t \sim \Lambda_{\rm UV}$ rather than its physical value.
The value of the top mass entering
the SM Higgs cross-section can easily be changed with a simple rescaling,
\beq
\label{eq:rescal_WC}
\sigma_H(m_S,\Gamma_S,\Lambda_{\rm UV}) = \left|\frac{ C(\mu_0, \Lambda_{\rm UV})}{ C(\mu_0,m_t)}\right|^2\,\sigma_H(m_S,\Gamma_S,m_t)\; .
\eeq

\begin{figure}[!t]
\begin{center}
\includegraphics[width=1.0\textwidth]{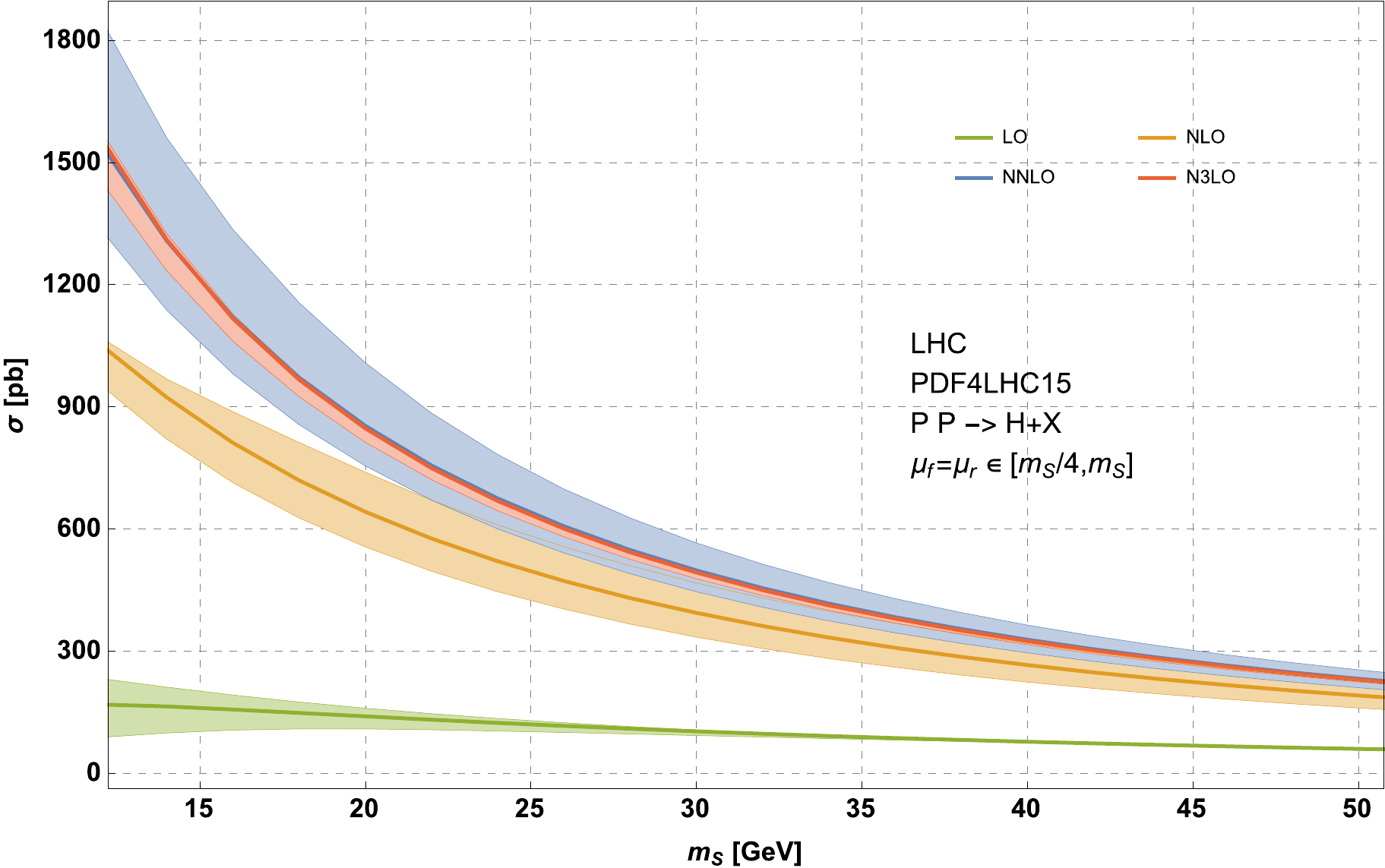}
\end{center}
\caption{
\label{fig:mhvarverylow}
Effective theory production cross section of a scalar particle as a function of the particle mass
$m_S\in\left[10,50 \right]$~GeV through increasing orders in perturbation
theory, at a 13~TeV proton-proton collider. The bands enveloping the respective orders represent the variation of the cross section due to variations of the scale $\mu\in \left[{m_S}/{4},m_S\right]$.
The value of the top mass is set to $m_t(m_t)=162.7$~GeV.
}
\end{figure}

\begin{figure}[!htb]
\begin{center}
\includegraphics[width=1.0\textwidth]{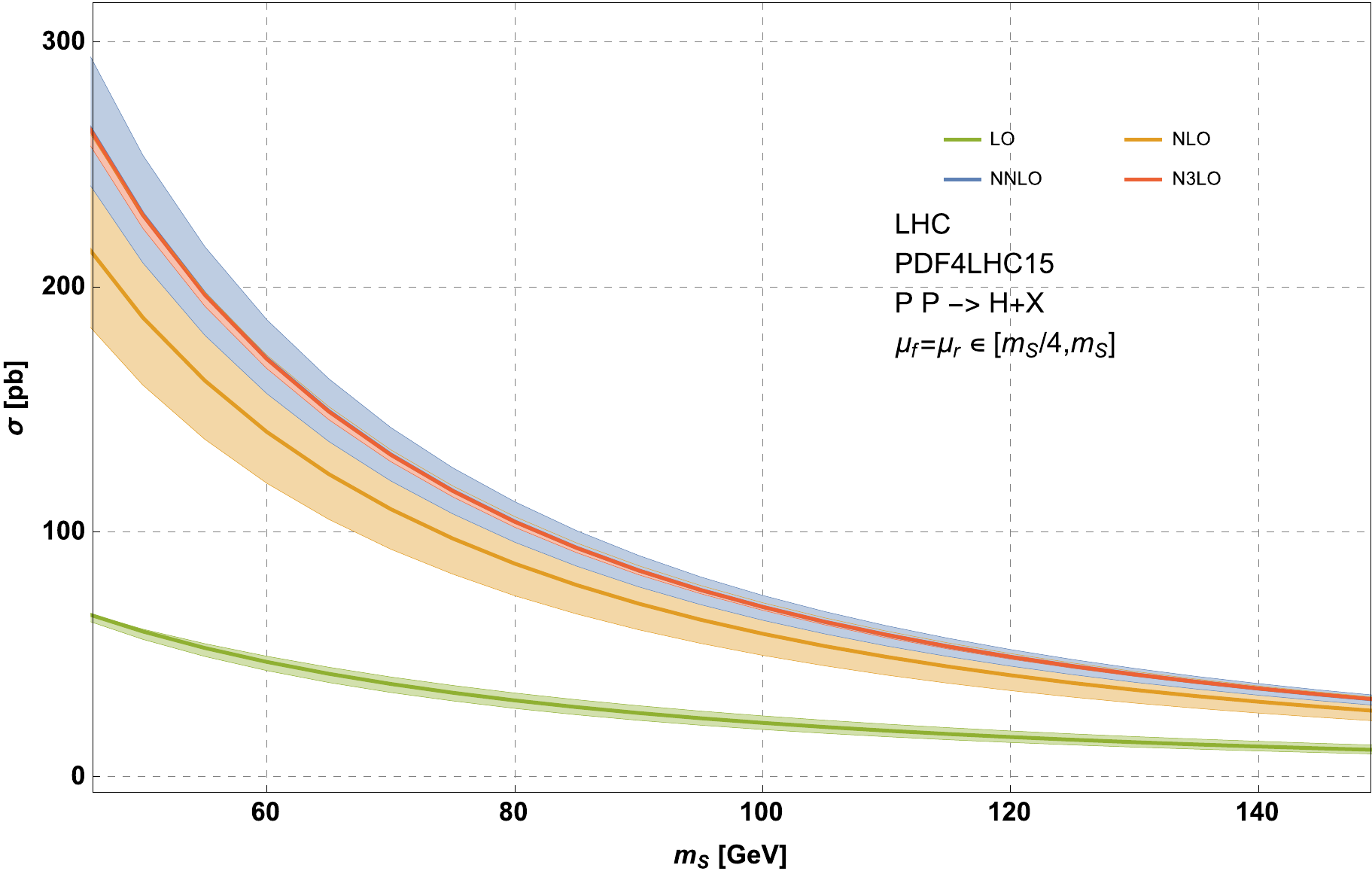}
\end{center}
\caption{
\label{fig:mhvarlow}
Effective theory production cross section of a scalar particle of mass
$m_S\in\left[50,150\right]$~GeV through increasing orders in perturbation
theory. For further details see the caption of Fig.~\ref{fig:mhvarverylow}.
}
\end{figure}

\begin{figure}[!htb]
\begin{center}
\includegraphics[width=1.0\textwidth]{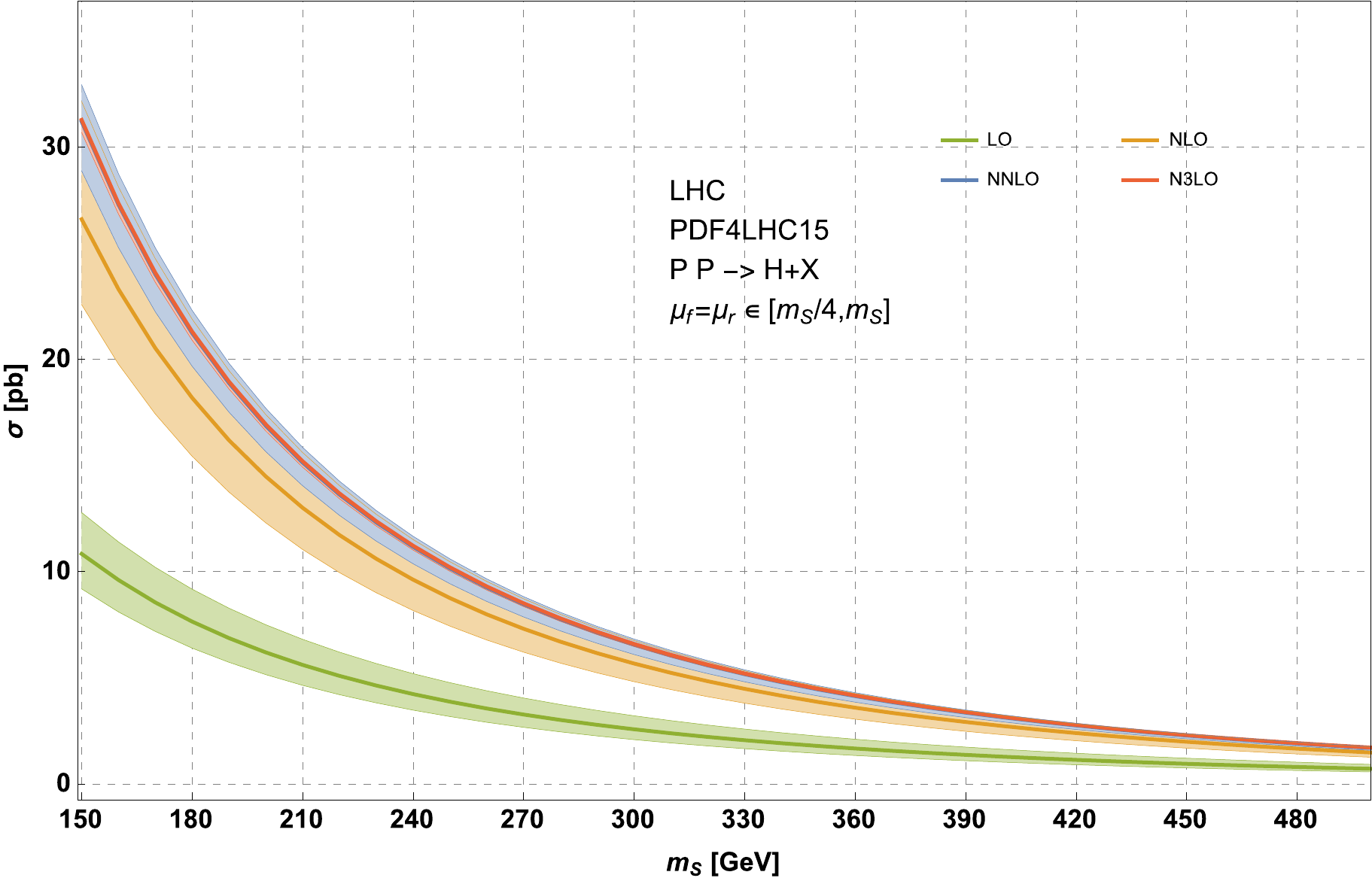}
\end{center}
\caption{
\label{fig:mhvarmid}
Effective theory production cross section of a scalar particle with mass
$m_S\in\left[150,500\right]$~GeV through increasing orders in
perturbation theory. For further details see the caption of Fig.~\ref{fig:mhvarverylow}.
}
\end{figure}
\begin{figure}[!htb]
\begin{center}
\includegraphics[width=1.0\textwidth]{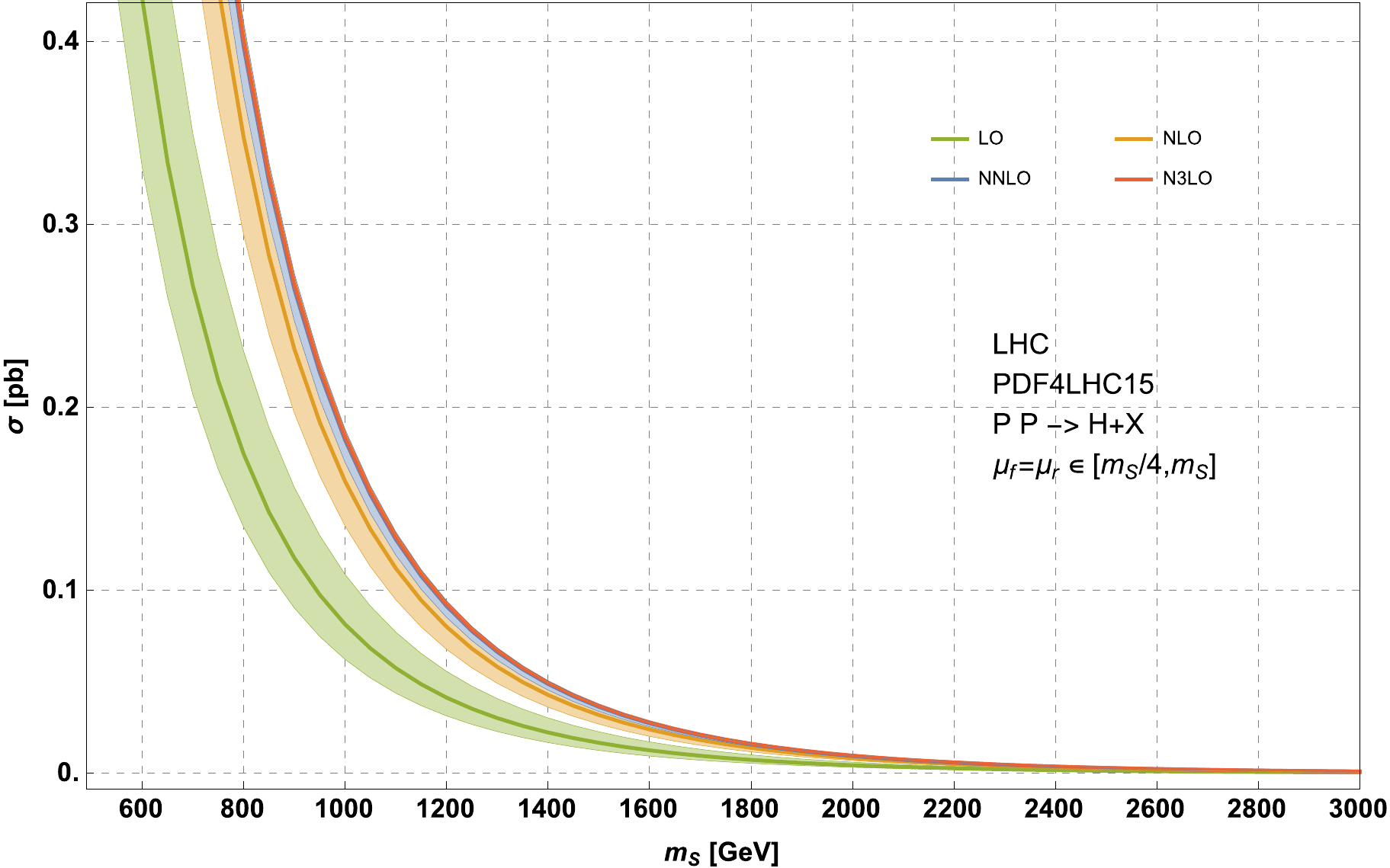}
\end{center}
\caption{
\label{fig:mhvarhigh}
Effective theory production cross section of a scalar particle as a function of the particle mass $m_S\in\left[500,3000\right]$~GeV through increasing orders in perturbation theory. For further details see the caption of Fig.~\ref{fig:mhvarverylow}.
}
\end{figure}

\begin{figure}[!htb]
\begin{center}
\includegraphics[width=1.0\textwidth]{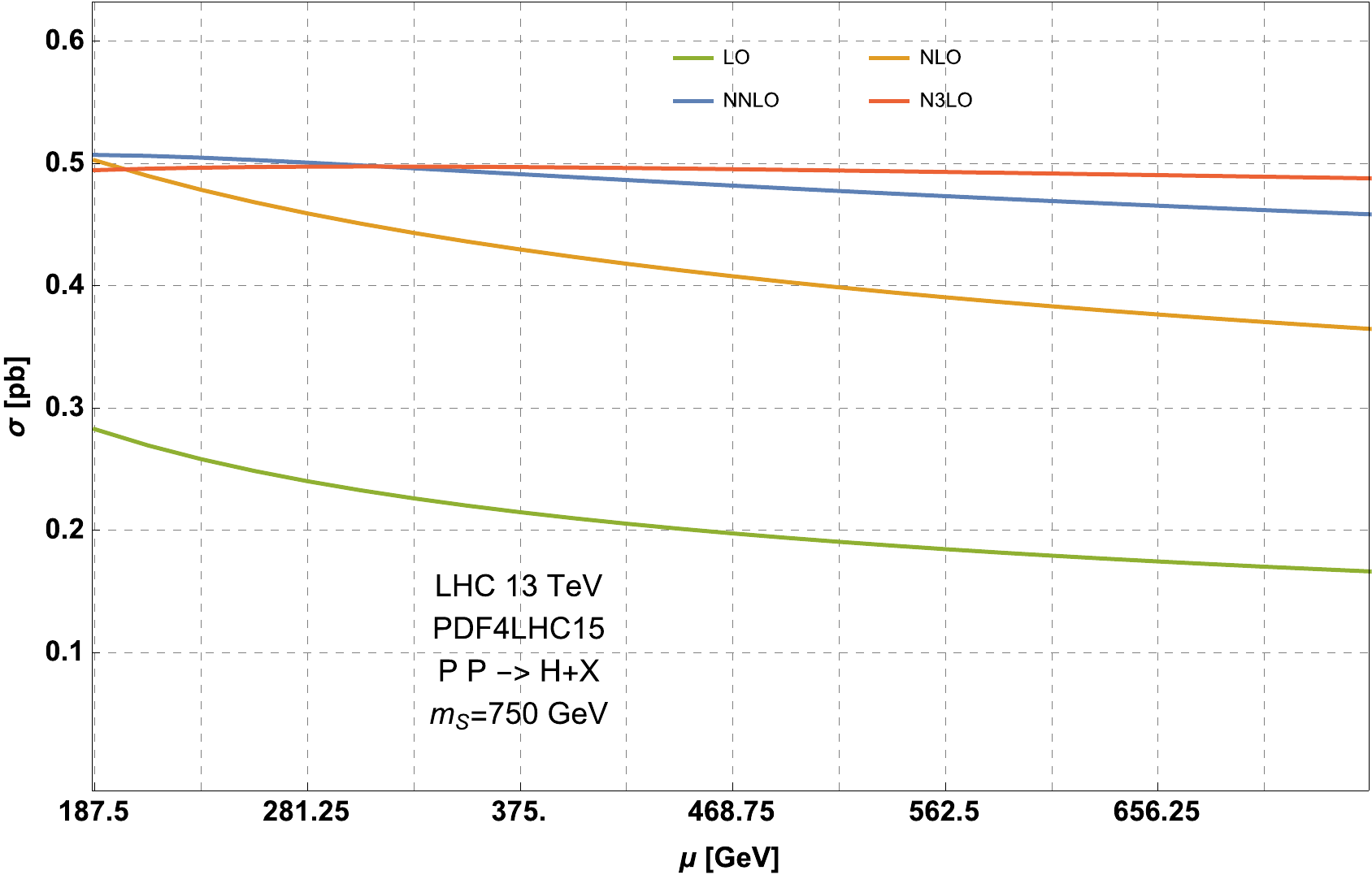}
\end{center}
\caption{
\label{fig:750var} Effective theory production cross section of a scalar particle with a mass
$m_S=$~750~GeV as a function of the common renormalization and factorization scales $\mu$
through increasing orders in perturbation theory.}
\end{figure}

Let us now discuss our predictions for the production cross section $\sigma_S$.
%In Figs.~\ref{fig:mhvarverylow},~\ref{fig:mhvarmid} and~\ref{fig:mhvarhigh}
In \mbox{Figs.~\ref{fig:mhvarverylow}\,-\,\ref{fig:mhvarhigh}}
we show the effective theory cross section
$\sigma_H(m_S,\Gamma_S=0,m_t)$ as a function of $m_S$
through N$^3$LO, at a proton-proton collider with a center-of-mass energy of 13~TeV.
The theory uncertainty bands correspond to a variation of the scale $\mu=\mu_r=\mu_f$ in the range
$\mu \in \left[{m_S}/{4}, m_S\right]$. The value of the top mass is set to $m_t(m_t)=162.7$~GeV
and we use the PDF4LHC15~\cite{Butterworth:2015oua} parton distribution function (PDF) set.
We observe that for the range of scalar masses between 10~GeV and 3~TeV
the \nnnlo scale variation band is always contained inside the NNLO band.
This is also true for the lowest values of this mass range,
$m_S\sim 10-50 \textrm{ GeV}$ (Fig.~\ref{fig:mhvarverylow}), where one observes especially
large corrections at NLO and NNLO. For higher masses, the scale variation is reduced.
As an illustration, we show in Fig.~\ref{fig:750var} the scale dependence for the cross section
of a CP-even scalar with a mass of 750~GeV.

Sources of theoretical uncertainty affecting the gluon-fusion
cross section at N$^3$LO
other than scale variation have been considered in detail in ref.~\cite{Anastasiou:2016cez}.
In particular, these sources of uncertainty are due to the lack of \nnnlo parton
densities and to the truncation of the threshold expansion for the \nnnlo correction.
In order to estimate the uncertainty on our computation,
we follow faithfully the uncertainty estimation prescription of ref.~\cite{Anastasiou:2016cez} { with one modification: the uncertainty due to the lack of \nnnlo parton densities is estimated here as the envelope of the corresponding uncertainties formed by using the CT14~\cite{Dulat:2015mca}, NNPDF30~\cite{Ball:2014uwa} and PDF4LHC sets, which are themselves computed according to  the prescription of ref.~\cite{Anastasiou:2016cez}. The reason for modifying the prescription in this case is that the PDF4LHC set leads to very small uncertainties for scalars in the region $500-1000$~GeV and, in particular, to a vanishing uncertainty estimate for a scalar with mass around 770~GeV. This feature is not shared with individual PDF sets. We therefore use, conservatively, the envelope of CT14, NNPDF30 and PDF4LHC, which leads to an uncertainty due to the lack of \nnnlo parton densities at the level of $0.9\%-3\%$ for scalars in the range $50$~GeV$-3$~TeV. This uncertainty remains of the order
of a few percent also at lower masses, but it increases rapidly to ${\cal O}(10\%)$ for 
$m_S \lesssim 20$~GeV.  }

We present the cross section values
and uncertainties for this range of scalar masses in Appendix~\ref{app:tables}.
In particular, in Tab.~\ref{tab:xsvals750} we focus on the range between 730 and 770~GeV.

%We should notice at this point that the effective theory approach only holds, strictly
%speaking, for $m_S\lesssim 2 m_t$. For higher values of the scalar mass,
%one should keep the full top-mass dependence into account, as described for
%example in ref.~\cite{Anastasiou:2016cez}.
%However, the use of the top mass as our reference value does not affect the
%correctness of our results. Following eq.~(\ref{eq:factorization}), one can combine the
%universal matrix element $\eta$ reported in the tables of Appendix~\ref{app:tables} with any arbitrary
%Wilson coefficient for the effective gluon-Higgs interaction where some heavy
%new physics has been integrated out to derive the \nnnlo EFT cross section
%in that specific model.

%%%%%%%%%%%%%%%%%%%
\section{Finite width effects and the line-shape}
\label{sec:width}
\begin{figure}[!t]
\begin{center}
\includegraphics[width=1.0\textwidth]{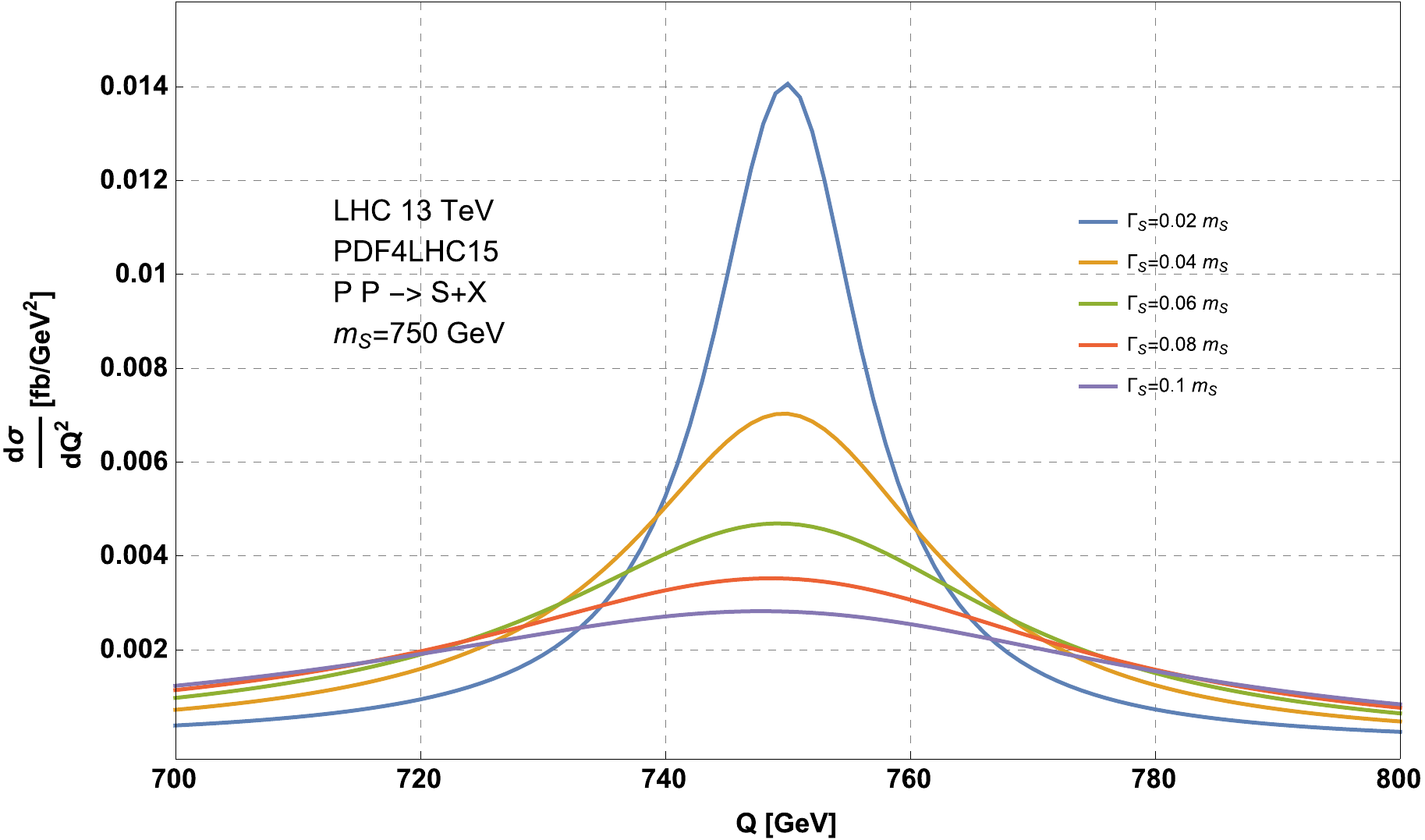}
\end{center}
\caption{
\label{fig:lineshape}
Line-shape of a 750~GeV CP-even scalar boson at the LHC for different
values of its total width.}
\end{figure}

The results of the previous section hold formally only when the width of the scalar is set to zero.
In many beyond the Standard Model (BSM) scenarios, however, finite-width effects
cannot be neglected. In this section
we present a way to include leading finite-width effects into our results,
in the case where the width is not too large compared to the mass.\\
The total cross section for the production of a scalar boson
%with arbitrary total width $\Gamma_S$
of total width $\Gamma_S$ can be obtained from the cross section in the
zero-width approximation via a convolution
\begin{equation}
\label{eq:width_convolution}
\sigma_S(m_S,\Gamma_S,\Lambda_{\rm UV})= \int dQ^2 \frac{Q \Gamma_S(Q)}{\pi}
\frac{\sigma_S(Q,\Gamma_S=0,\Lambda_{\rm UV})}{(Q^2-m_S^2)^2+ m_S^2 \Gamma^2(m_S)} + \mathcal{O}\left(\Gamma_S(m_S)/m_S\right)\,,
\end{equation}
where $Q$ is the virtuality of the scalar particle.
This expression
is accurate at leading order in $\Gamma_S(m_S)/m_S$. For
large values of the width relative to the mass, subleading corrections and
signal-background interference effects are important
%~\cite{Binoth:2006mf, Campbell:2011cu,
%Kauer:2012hd, Caola:2013yja}
and are not captured by eq.~\eqref{eq:width_convolution}. Let us also note that in order to
use eq.~\eqref{eq:width_convolution} faithfully, one needs the width
as a function of the virtuality of the scalar, which may bear a substantial model
dependence. However, it is often the case that the width can be
approximated as
\begin{equation}
\Gamma_S(Q \approx m_S) \equiv \Gamma_S\,.
\end{equation}
%Using this approximation, we plot in Fig.~\ref{fig:lineshape} the invariant mass
%distribution for the production of a CP-odd scalar with a mass of $m_S
%= 750$~GeV and total width from 2 to $10\%$ of the scalar mass, using as input an
%interpolation of the zero-width cross section values of
%Tab.~\ref{tab:xsvals750} in Appendix~\ref{app:tables}.
The invariant mass distribution in this approximation for the production of a
CP-even scalar with a mass of $m_S= 750$~GeV and total width from 2 to $10\%$
of the scalar mass is shown in Fig.~\ref{fig:lineshape}. This result has been
obtained from an
interpolation of the zero-width cross section values of
Tab.~\ref{tab:xsvals750} in Appendix~\ref{app:tables}.
We caution that if the results of
Appendix~\ref{app:tables}
are used to derive cross section values with non-zero width effects  following the strategy outlined in this section,
an additional
uncertainty of the order ${\cal O}\left({\Gamma_S}/{m_S}\right)$
should be assigned.

In order to facilitate the computation of the line-shape we perform a parametric fit of the
production cross section of a zero-width scalar boson as a function of
its virtuality (with $\Lambda_{\textrm{UV}}=m_S$) for a center-of-mass energy of $13$TeV. To guarantee agreement of the fitted cross section with the actual cross section at a level better than $1\%$, we split the range of interesting scalar boson mass values into three intervals:
\begin{itemize}
\item in the range $m_S\in [10\textrm{ GeV},150\textrm{ GeV}]$,
cf. Tab.~\ref{tab:xsvalslow}  in Appendix~\ref{app:tables}, we find
\bea
\label{eq:interpol1}
\sigma_S(x)&\approx & \left(3.89881\times 10^6\, x^2 -1.90274\times 10^6\, x -202261\,x\, \log ^2x+1623.77\, \log ^2x \right. \nonumber\\
&-&\left. 923052 \,x\, \log x+24108.2\, \log x+95652.2\right) \textrm{ pb}\,,
\eea
\item in the range $m_S\in [150\textrm{ GeV},500\textrm{ GeV}]$, cf. Tab.~\ref{tab:xsvalsmid}, we find
\beq
\label{eq:interpol2}
\sigma_S(x) \approx  \left(1-\sqrt[3]{x}\right)^{9.52798}\, x^{-0.0415044 \,\log x-1.50381}\,\, \textrm{ pb}\,,
\eeq
\item in the range $m_S\in [500\textrm{ GeV},3000\textrm{ GeV}]$, cf.
Tabs.~\ref{tab:xsvalshigh}-\ref{tab:xsvalshigh2}, we find
\beq
\label{eq:interpol3}
\sigma_S(x) \approx \left(1-\sqrt[3]{x}\right)^{9.71562} \,x^{-0.0040194\,\log^3x-0.0474683\, \log^2x-0.240878 \,\log x-1.81243} \textrm{ pb}\,,
\eeq
\end{itemize}
where $x \equiv \frac{Q/\textrm{GeV}}{13 \textrm{ TeV}}$. The fits of Eqs.~\eqref{eq:interpol1}-\eqref{eq:interpol3}
can be used as the kernel of the convolution in eq.~\eqref{eq:width_convolution}.

%%%%%%%%%%%%%%%%%%%
\section{Validity of the EFT approach}
\label{Sec:EFT_vs_exact}

So far, we have assumed that the effective theory of eq.~\eqref{eq:L_eff} furnishes an accurate
description of the gluon-scalar interaction.  However, this assumption may be challenged if the scalar
couples to light coloured particles. To investigate this effect, we will consider a scalar of mass
$m_S= 750$~GeV which couples to a  top-like quark of mass $m_T$.
In this scenario, we can compute the cross-section
exactly through NLO in perturbative QCD.  We can then compare this prediction with the cross-section derived
with the effective theory of eq.~\eqref{eq:L_eff}.

\begin{figure}[!t]
\begin{center}
\includegraphics[width=1.0\textwidth]{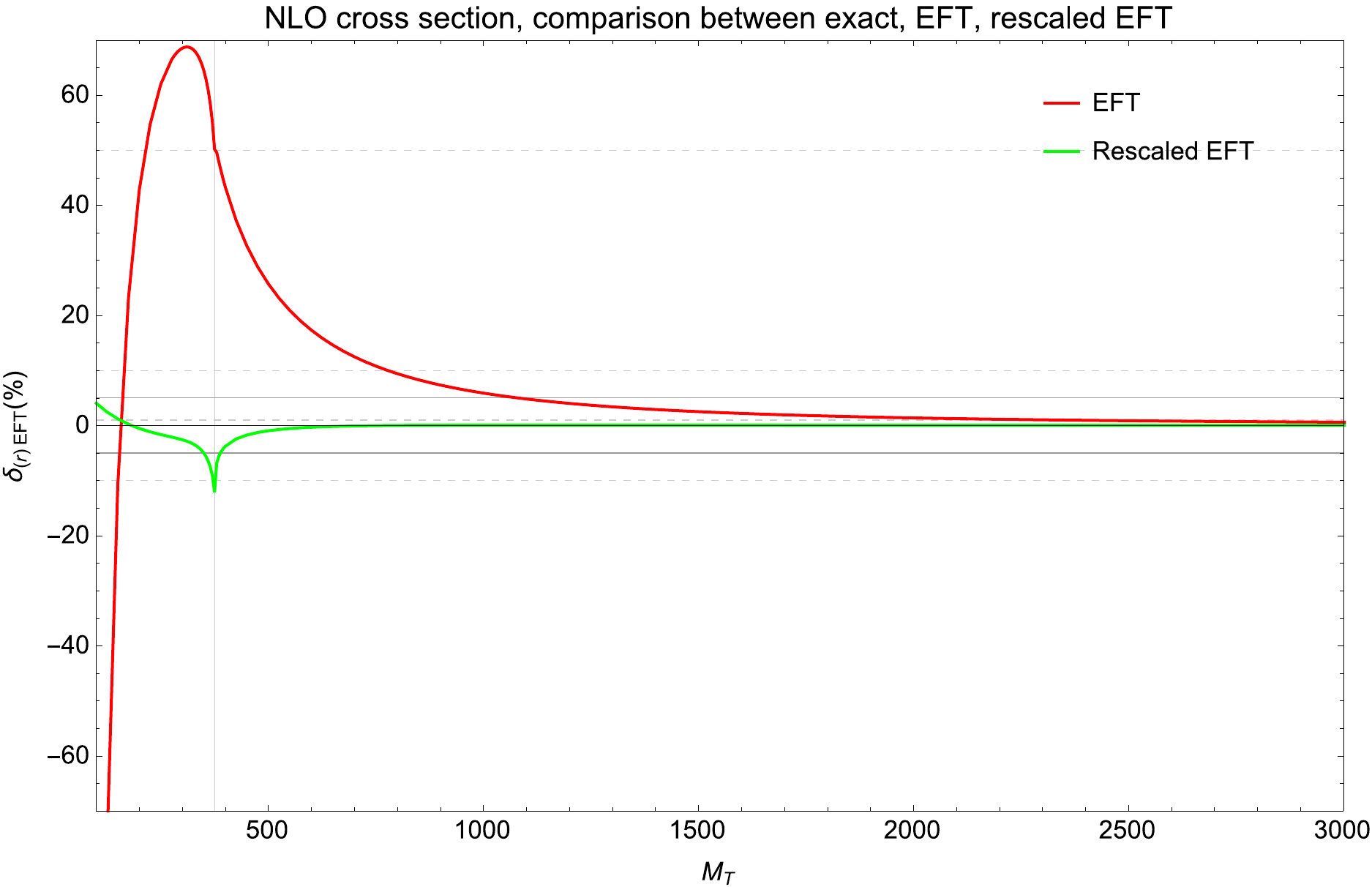}
\end{center}
\caption{
Percent difference~(\ref{eq:deltaEFT_ex}) between the exact and rescaled EFT (rEFT) cross section
at NLO (red line/green line) as a function of the quark mass for the production of a
750~GeV CP-even scalar. The vertical line corresponds to $m_S/2$. \newline
}
\label{Fig:(r)EFT_vs_Ex}
\end{figure}

The red line in Fig.~\ref{Fig:(r)EFT_vs_Ex} shows the percent difference of the two predictions,
\begin{equation}
\delta_{\rm EFT} = \frac{\sigma^{\rm NLO}_{\rm exact} -\sigma^{\rm
    NLO}_{\rm EFT}}{\sigma^{\rm NLO}_{\rm exact}} \times 100\% \; ,
\end{equation}
as a function of $m_T$ for a scalar with a mass of $m_S = 750$~GeV.
While in the region $m_T \sim  m_S/2$ the effective field theory
is inadequate, for larger values of $m_T$ the EFT description becomes accurate very quickly.
Indeed, for $m_T \sim 1.5 m_S$ the discrepancy with respect to the exact result is already below the
 theoretical uncertainty of QCD origin (of about $5\%$).

Let us now define a rescaled effective theory cross-section,
\begin{equation}
  \label{eq:rEFT_def}
  \sigma^{\rm NLO}_{\rm rEFT}= \frac{\sigma^{{\rm LO}}_{\rm exact}}{\sigma^{{\rm LO}}_{\rm EFT}}\,  \sigma^{{\rm NLO}}_{\rm EFT} \;.
\end{equation}
The green line of Fig.~\ref{Fig:(r)EFT_vs_Ex} shows the relative difference
\begin{equation}
	\label{eq:deltaEFT_ex}
	\delta_{\rm rEFT} = 100 \times \, \frac{\sigma^{\rm
            NLO}_{\rm exact}-\sigma^{\rm NLO}_{\rm rEFT}}{\sigma^{\rm NLO}_{\rm exact}}
\end{equation}
between the  rescaled effective theory prediction for the cross-section and the exact NLO
cross section.  We notice that the rescaled effective theory cross-section
reproduces the exact cross section very well (as it has also been
the case for the Standard Model Higgs boson~\cite{Harlander:2009mq,Pak:2009dg}).

We conclude that, while for the study of scalars which couple to relatively
light particles one should resort to a calculation within a
specific model, also in such situations the effective theory
computation can be utilised as a means to compute with a good accuracy
the corresponding QCD $K-$~factor with respect to the exact LO cross section.

\section{Top-quark contributions}
\label{sec:finite-top}
So far we have considered the case where the interaction between the heavy scalar $S$
and the Standard Model is mediated exclusively by the dimension-five operator in
eq.~\eqref{eq:L_eff}. In many extensions of the Standard Model
new scalars may also couple directly to the quarks of the third generation as well as
the $W$ and $Z$ bosons. However, the absence of resonances in top-pair
production and four-lepton production suggests that these couplings
should be small. It may be nonetheless useful to estimate the
contribution to the inclusive $S$-scalar cross section due to the heavy
SM quarks and gauge bosons for the purpose of setting precise experimental constraints
on their couplings to $S$. In this section, we study the effects due to the top quark, which
in many scenarios are expected to give the most important contribution.

For light scalar masses, $m_S < {\cal O} ( m_t)$,
contributions due to the coupling of the top quark can be simply taken
into account through \nnnlo in
perturbative QCD by using an appropriate Wilson coefficient $C_S$.
For heavier scalars with masses around and above the top-pair
threshold, however, it is not justified theoretically to integrate out the
top-quark. For this reason we study in the following the effect of modifying the
Largangian in eq.~\eqref{eq:L_eff} by including a direct Yukawa interaction
between the scalar $S$ and the top quark, i.e., we consider the Lagrangian
\beq\label{eq:L_eff_gen}
\mathcal{L}_{\text{eff}}=\mathcal{L}_{\textrm{SM}}+\mathcal{L}_S-\frac{\lambda_{\rm
    wc}}{4 v}C  \,
S\, G_{\mu\nu}^a G_a^{\mu\nu} - \lambda_t \frac{m_t}{v}\, S\, \bar t t\,,
\eeq
where $\lambda_t$ is the ratio
of the Yukawa coupling of the scalar $S$
and the Yukawa coupling of the Higgs boson to the top quark
and  $\lambda_{\rm wc}= \frac{C_S}{C}$ denotes the ratio of the Wilson coefficients
in this theory and the SM with the top-quark integrated out.

\begin{figure}[!t]
\begin{center}
\includegraphics[width=1.0\textwidth]{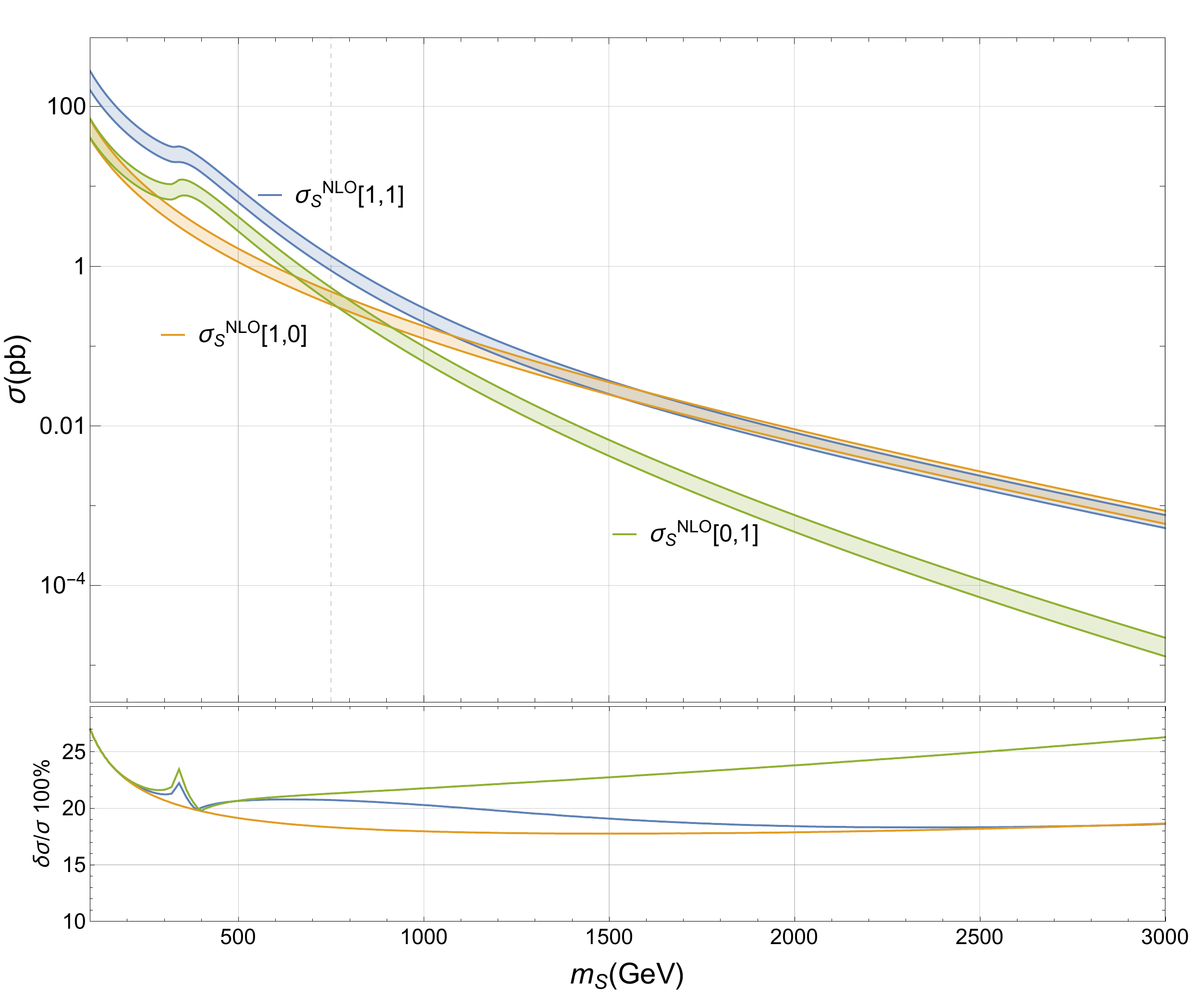}
\end{center}
\caption{
\label{fig:master_bsm_contributions}
NLO contributions to the total cross section for a scalar particle of mass $m_S$ as described in eqs.~\eqref{eq:master_bsm_unimproved}, \eqref{eq:master_bsm}. The bands correspond to the theory uncertainty (based on eq.~\eqref{eq:NLO_err_incl_sc}). In the lower pane we present the relative theory uncertainty ($\%$) for each of the contributions.
}
\end{figure}

The production cross section now depends on
the values of $\lambda_{\rm wc}$ and $\lambda_t$:
\begin{equation}
\sigma_S \equiv \sigma_S[\lambda_{\rm wc}, \lambda_t]\,.
\end{equation}
It does not take too much effort to show that the NLO cross section can be cast in the form
\beq\bsp
\label{eq:master_bsm_unimproved}
\sigma_S^{\rm NLO}[\lambda_{\rm wc}, \lambda_t] &\,=
\lambda_{\rm wc} (\lambda_{\rm wc} - \lambda_t )   \sigma_S^{\rm NLO}[1, 0]
+ \lambda_t (\lambda_t - \lambda_{\rm wc} ) \sigma_S^{\rm NLO}[0,
                1] \\
&\,+ \lambda_{\rm wc} \lambda_t \, \sigma_S^{\rm NLO}[1, 1]   \,.
\esp\eeq
The cross section with no Yukawa coupling, $\sigma_S^{\rm NLO}[1, 0]$, is known through N$^3$LO.
Hence, we can improve the previous expression by including all N$^3$LO corrections to the terms proportional to
$\lambda_{\rm wc}^2$,
\beq\bsp
\label{eq:master_bsm}
\sigma_S[\lambda_{\rm wc}, \lambda_t] &\,=
\lambda_{\rm wc}^2    \sigma_S^{\textrm{N}^3\textrm{LO}}[1,0]
-\lambda_{\rm wc} \lambda_t    \sigma_S^{\rm NLO}[1, 0]
+ \lambda_t (\lambda_t - \lambda_{\rm wc}) \sigma_S^{\rm NLO}[0,1]   \\
&\,+ \lambda_{\rm wc} \lambda_t \, \sigma_S^{\rm NLO}[1, 1]   \,.
\esp\eeq
\begin{table}[!t]
%\tiny\setlength{\tabcolsep}{5pt}
%
%    table 1
%
\centering
\label{tab:master_bsm_750}
\medskip
\begin{tabular}{c|c|c|c|c}
    \toprule
 $\sqrt{s}$ &Component &  value[fb] &    $\delta$(theory)\,[$\%$]    & $\delta$(pdf+$\alpha_S$)\,[$\%$]  \\
      \midrule
\multirow{4}{*}{7~TeV} &
 $\sigma_S^{\textrm{N}^3\textrm{LO}}[1,0]$ & 69.72 & ${}^{+2.0}_{-4.2}$  &  7.0\\
& $\sigma_S^{\rm NLO}[1,0]$  & 55.59 & 19.52 & 6.95   \\
& $\sigma_S^{\rm NLO}[0,1]$  & 61.71 & 22.69 & 6.94  \\
& $\sigma_S^{\rm NLO}[1,1]$  &152.6 & 22.1 & 6.92   \\
%\bottomrule
   \toprule
%% $\sqrt{s}$ &Component &  value[fb] &    $\delta$(theory)\,[$\%$]    & $\delta$(pdf+$\alpha_S$)\,[$\%$]  \\
%%      \midrule
\multirow{4}{*}{8~TeV} &
 $\sigma_S^{\textrm{N}^3\textrm{LO}}[1,0]$ & 111.4   &  ${}^{+1.9}_{-4.0}$     &  6.1 \\
& $\sigma_S^{\rm NLO}[1,0]$  & 89.37 & 19.18 & 6.23 \\
& $\sigma_S^{\rm NLO}[0,1]$  & 98.92 & 22.3 & 6.22 \\
& $\sigma_S^{\rm NLO}[1,1]$& 245.3 & 21.71 & 6.2 \\

   \toprule
\multirow{4}{*}{13~TeV} &
 $\sigma_S^{\textrm{N}^3\textrm{LO}}[1,0]$ & 496.9   & ${}^{+2.0}_{-3.7}$  & 4.0 \\
& $\sigma_S^{\rm NLO}[1,0]$  & 404.6  & 18.3  & 4.5 \\
& $\sigma_S^{\rm NLO}[0,1]$  & 442.7 & 21.3 & 4.4 \\
& $\sigma_S^{\rm NLO}[1,1]$  & 1108 & 20.7 & 4.4 \\
   \toprule
\multirow{4}{*}{14~TeV} &
 $\sigma_S^{\textrm{N}^3\textrm{LO}}[1,0]$ & 609.7 &  ${}^{+1.9}_{-3.7}$   & 3.8 \\
& $\sigma_S^{\rm NLO}[1,0]$  & 497.3 &18.21 & 4.26  \\
& $\sigma_S^{\rm NLO}[0,1]$  & 543. & 21.14 & 4.2 \\
& $\sigma_S^{\rm NLO}[1,1]$  & 1361   &  20.57  & 4.21 \\
\bottomrule
\end{tabular}
\caption{Contributions to the cross section~\eqref{eq:master_bsm} for the 
production of a CP-even scalar with mass 750~GeV at a proton-proton collider.}
\end{table}
Although we have taken into account QCD corrections within the EFT framework through \nnnlo,
the result for the cross section is formally only NLO-accurate because we are missing finite-mass effects beyond NLO.
We therefore estimate the uncertainties on the NLO cross-section
as
\begin{equation}
\label{eq:NLO_err_incl_sc}
\frac{	\delta \sigma^{\rm NLO}[n_1,n_2]}
{
 \sigma^{\rm NLO}[n_1,n_2]
} = \pm \delta_{>{\rm NLO}} \, (1 + \delta_{\rm scheme}[n_1,n_2])\,,\quad n_i\in\{0,1\}\,,
\end{equation}
where
\begin{equation}
\delta_{>{\rm NLO}}
= \left( \frac{\sigma^{\rm N^3LO}[1,0]-\sigma^{\rm NLO}[1,0]}
			{\sigma^{\rm NLO}[1,0]} \right)_{\rm EFT}
\end{equation}
is the relative change of the gluon-fusion cross-section in the
effective theory from NLO to N$^3$LO. Note that in this way make the assumption
that the cross-section components that are not known beyond NLO
will not have a worse perturbative convergence than in the
effective theory.  Finally,
we enlarge this uncertainty further by
\begin{equation}
\delta_{\rm scheme}[n_1,n_2]  =
			\frac{
\left|
\sigma^{\rm NLO, \overline{MS}}_{\textrm{exact}}[n_1,n_2] -\sigma^{\rm NLO,
  OS}_{\textrm{exact}}[n_1,n_2]
\right|
}
{\sigma^{\rm NLO, \overline{MS}}_{\textrm{exact}}[n_1,n_2]}\;,
\end{equation}
which measures the scheme dependence of the top-quark contribution at
NLO.

Equation~\eqref{eq:master_bsm} is our best prediction for the gluon-fusion production cross section of a generic scalar $S$.
We recall that the values of $\sigma_S^{\textrm{N}^3\textrm{LO}}[1,0]$ as a function of
the scalar boson mass can be read off from Tabs.~\ref{tab:xsvalslow}-\ref{tab:xsvalshigh2} in Appendix~\ref{app:tables}. The NLO cross sections $\sigma_S^{\rm NLO}[n_1,n_2]$
($n_1,n_2 \in \{0,1\}$) are reported in Tabs.~\ref{tab:xsterms}-\ref{tab:xsterms7}.
As an illustration, we present the complete set of terms
entering in eq.~\eqref{eq:master_bsm} for a scalar of mass $m_S=750\textrm{ GeV}$ in Tab.~\ref{tab:master_bsm_750}.

%, taking into account the progression of the QCD corrections based on the experience of the effective theory computation and the scheme dependence at NLO.
%{\bf [CD: Why do we highlight here that there are ``uncertainties on the NLO components''? IT is not just NLO, but out full set of uncertainty prescriptions, no?]}

The following fits are a good approximation to the central values of
the cross sections introduced in this section. The fits are functions of $x=\frac{m_S/\textrm{GeV}}{13 \textrm{TeV}}$ valid for a center-of-mass energy of $13 $TeV and PDF4LHC15 parton distribution functions.
\begin{itemize}
\item In the range $m_S\in [50\textrm{ GeV},150\textrm{ GeV}]$, we find
\begin{gather}
  \begin{aligned}
    \sigma_S^{\textrm{NLO}}[1,1]/\textrm{pb} &= -4.79097\times 10^9 x^2-\frac{454.08}{x^2}-\frac{48.2912 \log x}{x^2}+1.4105\times 10^9x\\
   &-\frac{1.22684\times 10^6}{x}+1.79376\times 10^7 \log ^2x+1.11478\times 10^9 x \log
   x\\
   &+1.59613\times 10^8 \log x-\frac{168047 \log x}{x}+4.39955\times 10^8\,,
  \end{aligned}
\end{gather}
\begin{gather}
\begin{align}
 \nonumber \sigma_S^{\textrm{NLO}}[1,0]/\textrm{pb} &= 2.71176\times 10^9 x^2+\frac{302.023}{x^2}+\frac{32.2322 \log x}{x^2}-8.07812\times 10^8 x\\
\nonumber   &+\frac{787193}{x}-1.11305\times 10^7 \log ^2x-6.62451\times 10^8 x \log x\\
   &-9.8134\times 10^7 \log x+\frac{108365 \log x}{x}-2.68924\times 10^8\,,
\end{align}
\end{gather}
\begin{gather}
  \begin{aligned}
    \sigma_S^{\textrm{NLO}}[0,1]/\textrm{pb} &= 910970 -3.05552\times 10^7 x^2+7.62007\times 10^6 x-\frac{544.886}{x}\\
    &+26407.1 \log^2x+3.86595\times 10^6 x \log x+290640 \log x\\
    &-\frac{58.3182 \log x}{x}\,.
  \end{aligned}
\end{gather}
% \begin{eqnarray}
% \sigma_S^{\rm NLO}[1,1] &=&
% 4.01444\times 10^8 x^2+\frac{5.5092}{x^2}+\frac{0.551495 \log x}{x^2}-1.0438\times 10^8 x
% \nonumber \\ &&
% +\frac{26567.8}{x}-608765\, \log ^2x-6.07953\times 10^7 x
%   \log x
% \nonumber \\ &&
% -6.02239\times 10^6 \log x+\frac{3385.43 \log x}{x}-1.7747\times 10^7\,,
% \end{eqnarray}

% \begin{eqnarray}
% \sigma_S^{\rm NLO}[1,0] &=&
% 1.22335\times 10^8 x^2+\frac{1.88731}{x^2}+\frac{0.189694 \log
%                             x}{x^2}-3.20541\times 10^7 x
% \nonumber \\ &&
% +\frac{8750.33}{x}-194884. \log ^2x-1.89402\times 10^7
%   x \log x
% \nonumber \\ &&
% -1.91627\times 10^6 \log x+\frac{1120.24 \log x}{x}-5.62489\times 10^6\,,
% \end{eqnarray}
% \begin{eqnarray}
% \sigma_S^{\rm NLO}[0,1] &=&
% 5350.55\, -940329. x^2+189222. x+\frac{13.3836}{x}-26.5798 \log ^2x
% \nonumber \\ &&
% +66398.6 x \log x+903.255 \log x+\frac{1.45725 \log x}{x}\,.
% \end{eqnarray}
\item In the range $m_S\in [150\textrm{ GeV},350\textrm{ GeV}]$, we find
\begin{gather}
  \begin{align}
\nonumber    \sigma_S^{\textrm{NLO}}[1,1]/\textrm{pb} &=5.78391\times 10^{11} x^3-6.04905\times 10^{12} x^3 \log x-3.97065\times 10^{12}
   x^2\\
\nonumber   &-1.60239\times 10^{12} x^2 \log x-2.28312\times 10^{11} x-\frac{7.35167\times
   10^6}{x}\\
   &+1.05353\times 10^8 \log ^2x-4.9669\times 10^{10} x \log x\\
\nonumber   &+6.11874\times 10^8 \log x-\frac{865250 \log x}{x}+5.46682\times 10^8\,,
  \end{align}
\end{gather}
\begin{gather}
  \begin{align}
   \nonumber \sigma_S^{\textrm{NLO}}[1,0]/\textrm{pb} &=-1.18403\times 10^{10} x^3+8.55496\times 10^{10} x^3 \log x+5.7597\times 10^{10}x^2\\
\nonumber   &+2.3659\times 10^{10} x^2 \log x+3.48892\times 10^9x+\frac{121488.}{x}\\
   &-1.66608\times 10^6 \log ^2x+7.67582\times 10^8 x \log x\\
\nonumber   &-9.42982\times 10^6 \log x+\frac{14369.3 \log x}{x}-7.62306\times 10^6\,,
  \end{align}
\end{gather}
\begin{gather}
  \begin{align}
\nonumber    \sigma_S^{\textrm{NLO}}[0,1]/\textrm{pb} &=3.58507\times 10^{11} x^3-3.67114\times 10^{12} x^3 \log x-2.41271\times 10^{12}x^2\\
\nonumber   &-9.74512\times 10^{11} x^2 \log x-1.3909\times 10^{11} x-\frac{4.49756\times
   10^6}{x}\\
   &+6.4297\times 10^7 \log ^2x-3.02763\times 10^{10} x \log x\\
\nonumber   &+3.72922\times10^8 \log x-\frac{529486 \log x}{x}+3.31551\times 10^8\,.
  \end{align}
\end{gather}
% \begin{eqnarray}
% \sigma_S^{\rm NLO}[1,1] &=&
% 1.36669\times 10^{12} x^3-1.0987\times 10^{13} x^3 \log
%                             x-7.33736\times 10^{12} x^2
% \nonumber \\ &&
% -2.99843\times 10^{12} x^2 \log x-4.37976\times 10^{11}
%   x-\frac{1.5073\times 10^7}{x}
% \nonumber \\ &&
% +2.0877\times 10^8 \log
%                 ^2x-9.60416\times 10^{10} x \log x
% \nonumber \\ &&
% +1.19222\times 10^9 \log x
% -\frac{1.78161\times 10^6
%   \log x}{x}
% \nonumber \\ &&
% +1.00752\times 10^9\,,
% \end{eqnarray}
% \begin{eqnarray}
% \sigma_S^{\rm NLO}[1,0] &=&
% 1.79548\times 10^{10} x^3-1.06198\times 10^{11} x^3 \log
%                             x-7.28076\times 10^{10}
%                             x^2
% \nonumber \\ &&
% -3.03204\times 10^{10} x^2 \log x
% -4.59222\times 10^9
%   x-\frac{172004.}{x}
% \nonumber \\ &&
% +2.27231\times 10^6 \log ^2x
% -1.01913\times 10^9
%                 x \log x
% \nonumber \\ &&
% +1.26194\times 10^7 \log x-\frac{20441.6 \log
%   x}{x}+9.51124\times 10^6\,,
% \end{eqnarray}
% \begin{eqnarray}
% \sigma_S^{\rm NLO}[0,1] &=&
% 1.36669\times 10^{12} x^3-1.0987\times 10^{13} x^3 \log
%                             x-7.33736\times 10^{12}
%                             x^2
% \nonumber \\ &&
% -2.99843\times 10^{12} x^2 \log x
% -4.37976\times 10^{11}
%   x-\frac{1.5073\times 10^7}{x}
% \nonumber \\ &&
% -9.60416\times 10^{10} x \log x
% +1.19222\times 10^9 \log x-\frac{1.78161\times 10^6
%   \log x}{x}
% \nonumber \\ &&
% +1.00752\times 10^9+2.0877\times 10^8 \log
%                 ^2x\,.
% \end{eqnarray}
\item In the range $m_S\in [350\textrm{ GeV},500\textrm{ GeV}]$, we find
\begin{gather}
  \begin{align}
\nonumber    \sigma_S^{\textrm{NLO}}[1,1]/\textrm{pb} &=-1.55048\times 10^{10} x^2+\frac{20602.3}{x^2}+4.12759\times 10^{10} x^2 \log x\\
\nonumber    &+3.27683\times 10^{10} x+\frac{3.90175\times 10^6}{x}-1.00487\times 10^8 \log^2x\\
    &-\frac{1.21312\times 10^6 \log ^2x}{x}+1.28157\times 10^{10} x \log x\\
\nonumber    &-1.37351\times 10^8 \log x-\frac{6.26022\times 10^6 \log x}{x}+8.75837\times 10^8\,,
  \end{align}
\end{gather}
\begin{gather}
  \begin{align}
\nonumber    \sigma_S^{\textrm{NLO}}[1,0]/\textrm{pb} &=-1.74667\times 10^9 x^2+\frac{2345.03}{x^2}+4.57918\times 10^9 x^2 \log x\\
    &+3.64833\times 10^9 x+\frac{450670}{x}-1.12934\times 10^7 \log ^2x\\
\nonumber    &-\frac{137023 \log^2x}{x}+1.43013\times 10^9 x \log x-1.53273\times 10^7 \log x\\
\nonumber    &-\frac{704422 \log x}{x}+9.83324\times 10^7\,,
  \end{align}
\end{gather}
\begin{gather}
  \begin{align}
\nonumber    \sigma_S^{\textrm{NLO}}[0,1]/\textrm{pb} &=1.59986\times 10^9 x^2-\frac{2171.87}{x^2}-4.2072\times 10^9 x^2 \log x\\
    &-3.34991\times10^9 x-\frac{417719}{x}+1.03789\times 10^7 \log ^2x\\
\nonumber    &+\frac{126226 \log^2x}{x}-1.3131\times 10^9 x \log x+1.40646\times 10^7 \log x\\
 \nonumber   &+\frac{647893 \log x}{x}-9.03229\times 10^7\,.
  \end{align}
\end{gather}
% \begin{eqnarray}
% \sigma_S^{\rm NLO}[1,1] &=&
% 6.32439\times 10^8 x^2-\frac{1707.23}{x^2}-3.58438\times 10^8 x^2 \log
%                             x-1.02539\times 10^8
%                             x
% \nonumber \\ &&
% +\frac{2.66971\times 10^7}{x}
% -4.88782\times 10^7 \log
%   ^2x+\frac{543909. \log ^2x}{x}
% \nonumber \\ &&
% -6.49864\times 10^8 x \log x
% -2.43821\times 10^8 \log x
% \nonumber \\ &&
% +\frac{7.24512\times 10^6 \log x}{x}-5.85045\times
%   10^8\,,
% \end{eqnarray}
% \begin{eqnarray}
% \sigma_S^{\rm NLO}[1,0] &=&
% 4.29431\times 10^7 x^2-\frac{86.2468}{x^2}-2.5163\times 10^7 x^2 \log
%                             x
% \nonumber \\ &&
% -9.65121\times 10^6 x+\frac{1.53812\times 10^6}{x}-2.99593\times 10^6 \log
%   ^2x
% \nonumber \\ &&
% +\frac{30547.7 \log ^2x}{x}-4.28611\times 10^7 x \log x
% \nonumber \\ &&
% -1.54223\times 10^7 \log x+\frac{412766 \log x}{x}
% \nonumber \\ &&
% -3.70278\times 10^7\,,
% \end{eqnarray}
% \begin{eqnarray}
% \sigma_S^{\rm NLO}[0,1] &=&
% 2.19221\times 10^8 x^2-\frac{668.091}{x^2}-1.22144\times 10^8 x^2 \log
%                             x
% \nonumber \\ &&
% -2.86532\times 10^7 x+\frac{9.95935\times 10^6}{x}-1.77715\times 10^7 \log
%   ^2x
% \nonumber \\ &&
% +\frac{204935. \log ^2x}{x}-2.28508\times 10^8 x \log
%                 x-8.74226\times 10^7 \log x
% \nonumber \\ &&
% +\frac{2.71476\times 10^6 \log x}{x}-2.09713\times
%   10^8\,.
% \end{eqnarray}
\item In the range $m_S\in [500\textrm{ GeV},1000\textrm{ GeV}]$, we find
\begin{gather}
  \begin{aligned}
    \sigma_S^{\textrm{NLO}}[1,1]/\textrm{pb} &=1.0459\times 10^8 x^2+\frac{478.474}{x^2}-7.72699\times 10^7 x^2 \log x\\
    &-8.14486\times10^7 x-\frac{2.50557\times 10^6}{x}-95661.5 \log ^2x\\
    &-\frac{71863.4 \log^2x}{x}-7.67177\times 10^7 x \log x\\
    &-1.27372\times 10^7 \log x-\frac{802665 \log x}{x}-3.08306\times 10^7\,,
  \end{aligned}
\end{gather}
\begin{gather}
  \begin{aligned}
    \sigma_S^{\textrm{NLO}}[1,0]/\textrm{pb} &=1.76181\times 10^7 x^2+\frac{81.1265}{x^2}-1.3039\times 10^7 x^2 \log x\\
    &-1.37328\times10^7 x-\frac{421944}{x}-16421.7 \log ^2x\\
    &-\frac{12116.3 \log ^2x}{x}-1.29224\times10^7 x \log x\\
    &-2.14537\times 10^6 \log x-\frac{135235 \log x}{x}-5.19164\times10^6\,,
  \end{aligned}
\end{gather}
\begin{gather}
  \begin{aligned}
    \sigma_S^{\textrm{NLO}}[0,1]/\textrm{pb} &=2.43245\times 10^7 x^2+\frac{110.92}{x^2}-1.78774\times 10^7 x^2 \log x\\
    &-1.88708\times10^7 x-\frac{585093}{x}-22465.3 \log ^2x\\
    &-\frac{16766.8 \log ^2x}{x}-1.78573\times10^7 x \log x\\
    &-2.97483\times 10^6 \log x-\frac{187382 \log x}{x}-7.19613\times10^6\,.
  \end{aligned}
\end{gather}
\item In the range $m_S\in [1000\textrm{ GeV},3000\textrm{ GeV}]$, we find
\begin{gather}
  \begin{aligned}
    \sigma_S^{\textrm{NLO}}[1,1]/\textrm{fb} &= x^{-0.407849 \log ^3x-0.841896 \log ^2x+77.9612 x \log x+14.6265 \log x+49.7825}\\
    &\times \left(1-\sqrt[3]{x}\right)^{-1.74158}\,,
  \end{aligned}
\end{gather}
\begin{gather}
  \begin{aligned}
    \sigma_S^{\textrm{NLO}}[1,0]/\textrm{fb} &= x^{-0.0682707 \log ^3x-0.812125 \log ^2x-2.68171 x \log x-3.91644 \log x-10.0547}\\
    &\times \left(1-\sqrt[3]{x}\right)^{7.95188}\,,
  \end{aligned}
\end{gather}
\begin{gather}
  \begin{aligned}
    \sigma_S^{\textrm{NLO}}[0,1]/\textrm{fb} &= x^{-2.40603 x+0.812624 \log ^2x+18.7175 x \log x+6.90869 \log x+12.7991}\\
    &\times \left(1-\sqrt[3]{x}\right)^{8.05859}\,.
  \end{aligned}
\end{gather}
% \begin{eqnarray}
% \sigma_S^{\rm NLO}[1,1] & =&
%  x^{-37.1877 x+3.68149 \log
%                              ^2x+85.1253 x \log x+33.1768 \log
%                              x+78.3655}\\
%                              &&\times \left(1-\sqrt[3]{x}\right)^{0.089424}\,,
% \nonumber
% \end{eqnarray}
% \begin{eqnarray}
% \sigma_S^{\rm NLO}[0,1] &=&
%  x^{-6.39169 x+0.0687294 \log
%                             ^2x+1.68005 x \log x+0.301381 \log
%                             x-2.4915}\\
%                             &&\times\left(1-\sqrt[3]{x}\right)^{7.78924}\,,
% \nonumber
% \end{eqnarray}
% \begin{eqnarray}
% \sigma_S^{\rm NLO}[1,0] &=&
%  x^{-5.29465 x+0.531992 \log
%   ^2(x)+9.62454 x \log x+4.23831 \log x+5.79517}\\
%   &&\times\left(1-\sqrt[3]{x}\right)^{10.0493}\,.
% \nonumber
% \end{eqnarray}
\end{itemize}
%In the above, $x \equiv \frac{m_S/\textrm{GeV}}{13 \textrm{ TeV}}$.

%%% Local Variables:
%%% mode: latex
%%% TeX-master: "paper"
%%% End:

%\input{Wilson.tex}
\section{Conclusion}
\label{sec:concl}
In this paper we have presented the most precise predictions for the production of a CP-even scalar produced in gluon fusion at the LHC. We assume that the coupling of the scalar to the gluons can be described in an effective field theory approach. This enables us to provide precision QCD corrections at N$^3$LO to the production cross section at the LHC. We find that for a scalar with a mass of $750$~GeV the N$^3$LO corrections yield a theoretical uncertainty of  $\sim 2\%$ when we choose the central renormalization and factorization scales to be $m_S/2$ and follow the prescriptions layed out in ref.~\cite{Anastasiou:2016cez}. 

If we assume that the effective theory Wilson coefficient coupling of the CP-even scalar and the gluons is generated by a heavy top-partner, we can estimate the validity of the effective description as a function of the heavy fermion mass. We find that in the threshold region, when the mass of the top-partner is about half the mass of the scalar, the corrections can be as large as $\sim60\%$. However, from direct searches the low mass ranges for top-partners are tightly constrained, so that at least for the discussion of a $750$~GeV scalar this configuration is unlikely. For heavier fermion masses the effective theory description performs very well and is already accurate at $\sim5\%$ when then fermion mass is about $~1.5$ times the scalar mass.

Additionally, we have considered the case that the new scalar couples directly to the SM top quark. In this case we need to take into account the top corrections to the cross section, which are unfortunately only available through NLO in QCD, and they thus come with uncertainties of up to $\sim20\%$. We incorporate the top corrections to the scalar production cross section as a function of the Yukawa coupling of the 
top quark to the scalar, allowing for a model-dependent choice of the parameters.

In summary, we have provided the ingredients to endow predictions for the production of a CP-even scalar at the LHC with the most precise 
QCD corrections available. The only assumptions that have been made in the calculation are that the scalar couples to the gluons through an 
effective theory like operator. By determining the Wilson coefficient for the scalar coupling to the gluons as well as the Yukawa coupling to the 
top within a concrete model, it is now possible, using the numbers provided here, to easily incorporate  QCD corrections through N$^3$LO into 
the cross section predictions for the LHC. 
While the results presented in this article concern strictly the production of a CP-even scalar, we can use the same techniques to compute the cross-section for resonance production of different CP/spin types which may be phenomenologically relevant  (see, for example, ref.~\cite{Panico:2016ary}).  This will be the subject of future works.

\section*{Acknowledgements}
CD, EF and TG are grateful to the KITP, Santa Barbara, for the hospitality during the 
final stages of this work.
We are grateful to Riccardo Barbieri and Giuliano Panico for useful dicussions. This research was supported  in part by the National Science Foundation under Grant No. NSF PHY11-25915, by the Swiss National Science Foundation (SNF) under  contracts 200021-165772 and  200020-162487 and by the European Commission 
through the ERC grants ``pertQCD'', ``HEPGAME'' (320651), ``HICCUP'', ``MathAm'' and ``MC@NNLO'' (340983).

%%% Local Variables:
%%% mode: latex
%%% TeX-master: "paper"
%%% End:

\bibliographystyle{JHEP.bst}
\bibliography{higgsrefs}

\providecommand{\href}[2]{#2}\begingroup\raggedright\begin{thebibliography}{10}

\bibitem{Aad:2012tfa}
{\scshape ATLAS} collaboration, G.~Aad et~al., \emph{{Observation of a new
  particle in the search for the Standard Model Higgs boson with the ATLAS
  detector at the LHC}},
  \href{http://dx.doi.org/10.1016/j.physletb.2012.08.020}{\emph{Phys. Lett.}
  {\bf B716} (2013) 1--29}, [\href{http://arxiv.org/abs/1207.7214}{{\tt
  1207.7214}}].

\bibitem{Chatrchyan:2012xdj}
{\scshape CMS} collaboration, S.~Chatrchyan et~al., \emph{{Observation of a new
  boson at a mass of 125 GeV with the CMS experiment at the LHC}},
  \href{http://dx.doi.org/10.1016/j.physletb.2012.08.021}{\emph{Phys. Lett.}
  {\bf B716} (2012) 30--61}, [\href{http://arxiv.org/abs/1207.7235}{{\tt
  1207.7235}}].

\bibitem{ATLAS_gaga}
{ATLAS collaboration}, \emph{{Search for resonances decaying to photon pairs in
  3.2 fb$^{-1}$ of $pp$ collisions at $\sqrt{s}$ = 13 TeV with the ATLAS
  detector}},  \href{http://arxiv.org/abs/ATLAS-CONF-2015-081}{{\tt
  ATLAS-CONF-2015-081}}.

\bibitem{CMS:2015dxe}
{\scshape CMS} collaboration, {CMS Collaboration}, \emph{{Search for new
  physics in high mass diphoton events in proton-proton collisions at 13TeV}},
  \href{http://arxiv.org/abs/CMS-PAS-EXO-15-004}{{\tt CMS-PAS-EXO-15-004}}.

\bibitem{Baikov:2009bg}
P.~A. Baikov, K.~G. Chetyrkin, A.~V. Smirnov, V.~A. Smirnov and M.~Steinhauser,
  \emph{{Quark and gluon form factors to three loops}},
  \href{http://dx.doi.org/10.1103/PhysRevLett.102.212002}{\emph{Phys. Rev.
  Lett.} {\bf 102} (2009) 212002}, [\href{http://arxiv.org/abs/0902.3519}{{\tt
  0902.3519}}].

\bibitem{Gehrmann:2010ue}
T.~Gehrmann, E.~W.~N. Glover, T.~Huber, N.~Ikizlerli and C.~Studerus,
  \emph{{Calculation of the quark and gluon form factors to three loops in
  QCD}}, \href{http://dx.doi.org/10.1007/JHEP06(2010)094}{\emph{JHEP} {\bf 06}
  (2010) 094}, [\href{http://arxiv.org/abs/1004.3653}{{\tt 1004.3653}}].

\bibitem{Hoschele:2012xc}
M.~H{\"o}schele, J.~Hoff, A.~Pak, M.~Steinhauser and T.~Ueda, \emph{{Higgs
  boson production at the LHC: NNLO partonic cross sections through order
  $\epsilon$ and convolutions with splitting functions to N$^3$LO}},
  \href{http://dx.doi.org/10.1016/j.physletb.2013.03.003}{\emph{Phys. Lett.}
  {\bf B721} (2013) 244--251}, [\href{http://arxiv.org/abs/1211.6559}{{\tt
  1211.6559}}].

\bibitem{Anastasiou:2012kq}
C.~Anastasiou, S.~Buehler, C.~Duhr and F.~Herzog, \emph{{NNLO phase space
  master integrals for two-to-one inclusive cross sections in dimensional
  regularization}},
  \href{http://dx.doi.org/10.1007/JHEP11(2012)062}{\emph{JHEP} {\bf 11} (2012)
  062}, [\href{http://arxiv.org/abs/1208.3130}{{\tt 1208.3130}}].

\bibitem{Anastasiou:2013srw}
C.~Anastasiou, C.~Duhr, F.~Dulat and B.~Mistlberger, \emph{{Soft triple-real
  radiation for Higgs production at N3LO}},
  \href{http://dx.doi.org/10.1007/JHEP07(2013)003}{\emph{JHEP} {\bf 1307}
  (2013) 003}, [\href{http://arxiv.org/abs/1302.4379}{{\tt 1302.4379}}].

\bibitem{Anastasiou:2013mca}
C.~Anastasiou, C.~Duhr, F.~Dulat, F.~Herzog and B.~Mistlberger,
  \emph{{Real-virtual contributions to the inclusive Higgs cross-section at
  $N^3LO$}}, \href{http://dx.doi.org/10.1007/JHEP12(2013)088}{\emph{JHEP} {\bf
  12} (2013) 088}, [\href{http://arxiv.org/abs/1311.1425}{{\tt 1311.1425}}].

\bibitem{Li:2013lsa}
Y.~Li and H.~X. Zhu, \emph{{Single soft gluon emission at two loops}},
  \href{http://dx.doi.org/10.1007/JHEP11(2013)080}{\emph{JHEP} {\bf 11} (2013)
  080}, [\href{http://arxiv.org/abs/1309.4391}{{\tt 1309.4391}}].

\bibitem{Anastasiou:2014vaa}
C.~Anastasiou, C.~Duhr, F.~Dulat, E.~Furlan, T.~Gehrmann, F.~Herzog et~al.,
  \emph{{Higgs boson gluon--fusion production at threshold in N$^3$LO QCD}},
  \href{http://dx.doi.org/10.1016/j.physletb.2014.08.067}{\emph{Phys. Lett.}
  {\bf B737} (2014) 325--328}, [\href{http://arxiv.org/abs/1403.4616}{{\tt
  1403.4616}}].

\bibitem{Anastasiou:2014lda}
C.~Anastasiou, C.~Duhr, F.~Dulat, E.~Furlan, T.~Gehrmann, F.~Herzog et~al.,
  \emph{{Higgs boson gluon-fusion production beyond threshold in N$^{3}$LO
  QCD}}, \href{http://dx.doi.org/10.1007/JHEP03(2015)091}{\emph{JHEP} {\bf 03}
  (2015) 091}, [\href{http://arxiv.org/abs/1411.3584}{{\tt 1411.3584}}].

\bibitem{Dulat:2014mda}
F.~Dulat and B.~Mistlberger, \emph{{Real-Virtual-Virtual contributions to the
  inclusive Higgs cross section at N3LO}},
  \href{http://arxiv.org/abs/1411.3586}{{\tt 1411.3586}}.

\bibitem{Duhr:2014nda}
C.~Duhr, T.~Gehrmann and M.~Jaquier, \emph{{Two-loop splitting amplitudes and
  the single-real contribution to inclusive Higgs production at N$^3$LO}},
  \href{http://dx.doi.org/10.1007/JHEP02(2015)077}{\emph{JHEP} {\bf 02} (2015)
  077}, [\href{http://arxiv.org/abs/1411.3587}{{\tt 1411.3587}}].

\bibitem{Li:2014afw}
Y.~Li, A.~von Manteuffel, R.~M. Schabinger and H.~X. Zhu, \emph{{Soft-virtual
  corrections to Higgs production at N$^3$LO}},
  \href{http://dx.doi.org/10.1103/PhysRevD.91.036008}{\emph{Phys. Rev.} {\bf
  D91} (2015) 036008}, [\href{http://arxiv.org/abs/1412.2771}{{\tt
  1412.2771}}].

\bibitem{Anastasiou:2015ema}
C.~Anastasiou, C.~Duhr, F.~Dulat, F.~Herzog and B.~Mistlberger, \emph{{Higgs
  Boson Gluon-Fusion Production in QCD at Three Loops}},
  \href{http://dx.doi.org/10.1103/PhysRevLett.114.212001}{\emph{Phys. Rev.
  Lett.} {\bf 114} (2015) 212001}, [\href{http://arxiv.org/abs/1503.06056}{{\tt
  1503.06056}}].

\bibitem{Anzai:2015wma}
C.~Anzai, A.~Hasselhuhn, M.~H{\"o}schele, J.~Hoff, W.~Kilgore, M.~Steinhauser
  et~al., \emph{{Exact N$^{3}$LO results for qq$^{\prime}\to$ H + X}},
  \href{http://dx.doi.org/10.1007/JHEP07(2015)140}{\emph{JHEP} {\bf 07} (2015)
  140}, [\href{http://arxiv.org/abs/1506.02674}{{\tt 1506.02674}}].

\bibitem{Anastasiou:2016cez}
C.~Anastasiou, C.~Duhr, F.~Dulat, E.~Furlan, T.~Gehrmann, F.~Herzog et~al.,
  \emph{{High precision determination of the gluon fusion Higgs boson
  cross-section at the LHC}},
  \href{http://dx.doi.org/10.1007/JHEP05(2016)058}{\emph{JHEP} {\bf 05} (2016)
  101}, [\href{http://arxiv.org/abs/1602.00695}{{\tt 1602.00695}}].

\bibitem{Anastasiou:2015yha}
C.~Anastasiou, C.~Duhr, F.~Dulat, E.~Furlan, F.~Herzog and B.~Mistlberger,
  \emph{{Soft expansion of double-real-virtual corrections to Higgs production
  at N$^{3}$LO}}, \href{http://dx.doi.org/10.1007/JHEP08(2015)051}{\emph{JHEP}
  {\bf 08} (2015) 051}, [\href{http://arxiv.org/abs/1505.04110}{{\tt
  1505.04110}}].

\bibitem{HXSWG}
{Higgs Cross-Section Working Group}, \emph{{BSM Higgs production cross sections
  at $\sqrt{s}$ = 13 TeV (update in CERN Report4 2016)}}, .

\bibitem{Dolan:2016eki}
M.~J. Dolan, J.~L. Hewett, M.~Kr{\"a}mer and T.~G. Rizzo, \emph{{Simplified
  Models for Higgs Physics: Singlet Scalar and Vector-like Quark
  Phenomenology}},  \href{http://arxiv.org/abs/1601.07208}{{\tt 1601.07208}}.

\bibitem{PDG}
{\scshape Particle Data Group} collaboration, S.~Eidelman et~al., \emph{{Review
  of particle physics. Particle Data Group}},
  \href{http://dx.doi.org/10.1016/j.physletb.2004.06.001}{\emph{Phys. Lett.}
  {\bf B592} (2004) 1--1109}.

\bibitem{Spiridonov:1988md}
V.~P. Spiridonov and K.~G. Chetyrkin, \emph{{Nonleading mass corrections and
  renormalization of the operators m psi-bar psi and g**2(mu nu)}}, {\emph{Sov.
  J. Nucl. Phys.} {\bf 47} (1988) 522--527}.

\bibitem{Chetyrkin:1997un}
K.~G. Chetyrkin, B.~A. Kniehl and M.~Steinhauser, \emph{{Decoupling relations
  to O (alpha-s**3) and their connection to low-energy theorems}},
  \href{http://dx.doi.org/10.1016/S0550-3213(97)00649-4}{\emph{Nucl. Phys.}
  {\bf B510} (1998) 61--87}, [\href{http://arxiv.org/abs/hep-ph/9708255}{{\tt
  hep-ph/9708255}}].

\bibitem{Schroder:2005hy}
Y.~Schroder and M.~Steinhauser, \emph{{Four-loop decoupling relations for the
  strong coupling}},
  \href{http://dx.doi.org/10.1088/1126-6708/2006/01/051}{\emph{JHEP} {\bf 01}
  (2006) 051}, [\href{http://arxiv.org/abs/hep-ph/0512058}{{\tt
  hep-ph/0512058}}].

\bibitem{Butterworth:2015oua}
J.~Butterworth et~al., \emph{{PDF4LHC recommendations for LHC Run II}},
  \href{http://dx.doi.org/10.1088/0954-3899/43/2/023001}{\emph{J. Phys.} {\bf
  G43} (2016) 023001}, [\href{http://arxiv.org/abs/1510.03865}{{\tt
  1510.03865}}].

\bibitem{Dulat:2015mca}
S.~Dulat, T.~J. Hou, J.~Gao, M.~Guzzi, J.~Huston, P.~Nadolsky et~al.,
  \emph{{The CT14 Global Analysis of Quantum Chromodynamics}},
  \href{http://arxiv.org/abs/1506.07443}{{\tt 1506.07443}}.

\bibitem{Ball:2014uwa}
{\scshape NNPDF} collaboration, R.~D. Ball et~al., \emph{{Parton distributions
  for the LHC Run II}},
  \href{http://dx.doi.org/10.1007/JHEP04(2015)040}{\emph{JHEP} {\bf 04} (2015)
  040}, [\href{http://arxiv.org/abs/1410.8849}{{\tt 1410.8849}}].

\bibitem{Harlander:2009mq}
R.~V. Harlander and K.~J. Ozeren, \emph{{Finite top mass effects for hadronic
  Higgs production at next-to-next-to-leading order}},
  \href{http://dx.doi.org/10.1088/1126-6708/2009/11/088}{\emph{JHEP} {\bf 11}
  (2009) 088}, [\href{http://arxiv.org/abs/0909.3420}{{\tt 0909.3420}}].

\bibitem{Pak:2009dg}
A.~Pak, M.~Rogal and M.~Steinhauser, \emph{{Finite top quark mass effects in
  NNLO Higgs boson production at LHC}},
  \href{http://dx.doi.org/10.1007/JHEP02(2010)025}{\emph{JHEP} {\bf 02} (2010)
  025}, [\href{http://arxiv.org/abs/0911.4662}{{\tt 0911.4662}}].

\bibitem{Panico:2016ary}
G.~Panico, L.~Vecchi and A.~Wulzer, \emph{{Resonant Diphoton Phenomenology
  Simplified}},  \href{http://arxiv.org/abs/1603.04248}{{\tt 1603.04248}}.

\end{thebibliography}\endgroup

\appendix
\section{Reference tables for the production cross section of a CP-even scalar through top-quark loops}
\label{app:tables}

Here we report the values of the \nnnlo production cross section
$\sigma_H(m_S, \Gamma_S=0, m_t)$ for a scalar of mass
$m_S \in [10\textrm{ GeV},3\textrm{ TeV}]$ at the 13~TeV LHC, in an effective
theory where the top quark has been integrated out. The $\overline{\textrm{MS}}$-mass of the top quark is chosen to be 162.7~GeV.
 We also show the two components
that enter this result (cf. eq.~(\ref{eq:master_bsm}), with
$\sigma_S \rightarrow\sigma_H$,
$\Lambda_{\rm UV} \rightarrow m_t$,
$C_S(\mu, \Lambda_{\rm UV}) \rightarrow C(\mu_0, m_t) $), i.e. the squared SM Wilson
coefficient $\left| C(\mu_0 = m_S/2, m_t)\right|^2$ and the matrix-element $\eta$. The theory error
is computed from the variation of the common renormalization and factorization scale $\mu$
in the range $\left[{m_S}/{4}, m_S \right]$. To it we add linearly the uncertainty from
missing \nnnlo PDFs, evaluated following the method of sec.~\ref{Sec:N3_EFT}, 
and from the truncation of the threshold expansion.
We refer to ref.~\cite{Anastasiou:2016cez} for an explanation of how the latter is computed.
We use the PDF set PDF4LHC14~\cite{Butterworth:2015oua}
and derive the combined PDF+$\alpha_s$ error following the indications of the PDF4LHC
working group.

\begin{table}[!h]
\begin{equation}
\begin{array}{|c|c|c|c|c|c|c|}
\hline
m_S \text{ [GeV]}&\hspace{0.2cm} C^2(\mu_0^2) \times 10^{-5}& \hspace{0.2cm} \sigma\text{ [pb]} \hspace{0.2cm}  & \hspace{0.2cm} \delta(\text{theory})\,[\%] \hspace{0.2cm}&  \hspace{0.2cm} \delta(\alpha_S+\text{PDF})\,[\%]    \\
 \hline 10 & 4.646 & 1899.6 & {}^{+17.8}_{-21.3} & \pm 12.2   \\ 
 \hline 15 & 3.806 & 1202.7 & {}^{+11.6}_{-15.} & \pm 6.7   \\ 
 \hline 20 & 3.348 & 845.8 & {}^{+8.8}_{-11.8} & \pm 5.4   \\ 
 \hline 25 & 3.05 & 632.2 & {}^{+7.1}_{-10.} & \pm 4.9   \\ 
 \hline 30 & 2.836 & 492.3 & {}^{+6.1}_{-8.8} & \pm 4.6   \\ 
 \hline 35 & 2.673 & 394.9 & {}^{+5.5}_{-7.8} & \pm 4.3   \\ 
 \hline 40 & 2.543 & 324. & {}^{+4.8}_{-7.3} & \pm 4.2   \\ 
 \hline 45 & 2.436 & 270.6 & {}^{+4.3}_{-6.7} & \pm 4.   \\ 
 \hline 50 & 2.347 & 229.4 & {}^{+3.7}_{-6.} & \pm 4.   \\ 
 \hline 55 & 2.27 & 196.8 & {}^{+3.5}_{-5.7} & \pm 3.8   \\ 
 \hline 60 & 2.203 & 170.6 & {}^{+3.2}_{-5.4} & \pm 3.7   \\ 
 \hline 65 & 2.144 & 149.2 & {}^{+3.1}_{-5.2} & \pm 3.7   \\ 
 \hline 70 & 2.092 & 131.5 & {}^{+2.8}_{-4.9} & \pm 3.6   \\ 
 \hline 75 & 2.045 & 116.6 & {}^{+2.6}_{-4.7} & \pm 3.6   \\ 
 \hline 80 & 2.002 & 104.1 & {}^{+2.5}_{-4.7} & \pm 3.5   \\ 
 \hline 85 & 1.964 & 93.4 & {}^{+2.4}_{-4.5} & \pm 3.5   \\ 
 \hline 90 & 1.928 & 84.2 & {}^{+2.3}_{-4.3} & \pm 3.4   \\ 
 \hline 95 & 1.896 & 76.3 & {}^{+2.2}_{-4.1} & \pm 3.4   \\ 
 \hline 100 & 1.865 & 69.3 & {}^{+2.}_{-4.} & \pm 3.4   \\ 
 \hline 105 & 1.837 & 63.2 & {}^{+2.}_{-3.9} & \pm 3.3   \\ 
 \hline 110 & 1.811 & 57.9 & {}^{+1.9}_{-3.9} & \pm 3.3   \\ 
 \hline 115 & 1.787 & 53.1 & {}^{+1.8}_{-3.8} & \pm 3.3   \\ 
 \hline 120 & 1.764 & 48.9 & {}^{+1.8}_{-3.7} & \pm 3.2   \\ 
 \hline 125 & 1.742 & 45.2 & {}^{+1.7}_{-3.7} & \pm 3.2   \\ 
 \hline 130 & 1.722 & 41.8 & {}^{+1.7}_{-3.6} & \pm 3.2   \\ 
 \hline 135 & 1.702 & 38.8 & {}^{+1.6}_{-3.5} & \pm 3.2   \\ 
 \hline 140 & 1.684 & 36. & {}^{+1.6}_{-3.5} & \pm 3.2   \\ 
 \hline 145 & 1.667 & 33.5 & {}^{+1.5}_{-3.5} & \pm 3.2   \\ 
 \hline 150 & 1.65 & 31.3 & {}^{+1.5}_{-3.4} & \pm 3.1   \\ 
\hline
\end{array}\nonumber
\end{equation}
\caption{\label{tab:xsvalslow} Production cross section for a scalar particle with a mass in the range 
10~GeV to 150~GeV. }
\end{table}

\begin{table}[!h]
\begin{equation}
\begin{array}{|c|c|c|c|c|c|c|}
\hline
m_S \text{ [GeV]}&\hspace{0.2cm} C^2(\mu_0^2) \times 10^{-5}& \hspace{0.2cm} \sigma\text{ [pb]} \hspace{0.2cm}  & \hspace{0.2cm} \delta(\text{theory})\,[\%] \hspace{0.2cm}&  \hspace{0.2cm} \delta(\alpha_S+\text{PDF})\,[\%]   \\
 \hline 150 & 1.65 & 31.29 & {}^{+1.5}_{-3.4} & \pm 3.1   \\
 \hline 160 & 1.619 & 27.37 & {}^{+1.5}_{-3.3} & \pm 3.1   \\
 \hline 170 & 1.591 & 24.09 & {}^{+1.4}_{-3.2} & \pm 3.1   \\
 \hline 180 & 1.565 & 21.32 & {}^{+1.3}_{-3.2} & \pm 3.1   \\
 \hline 190 & 1.542 & 18.96 & {}^{+1.3}_{-3.1} & \pm 3.   \\
 \hline 200 & 1.519 & 16.94 & {}^{+1.3}_{-3.2} & \pm 3.   \\
 \hline 210 & 1.499 & 15.2 & {}^{+1.4}_{-3.2} & \pm 3.   \\
 \hline 220 & 1.48 & 13.69 & {}^{+1.4}_{-3.2} & \pm 3.   \\
 \hline 230 & 1.461 & 12.37 & {}^{+1.4}_{-3.2} & \pm 3.   \\
 \hline 240 & 1.445 & 11.22 & {}^{+1.4}_{-3.2} & \pm 3.   \\
 \hline 250 & 1.428 & 10.2 & {}^{+1.4}_{-3.2} & \pm 3.   \\
 \hline 260 & 1.413 & 9.3 & {}^{+1.4}_{-3.2} & \pm 3.   \\
 \hline 270 & 1.399 & 8.51 & {}^{+1.4}_{-3.2} & \pm 3.   \\
 \hline 280 & 1.385 & 7.8 & {}^{+1.4}_{-3.2} & \pm 3.   \\
 \hline 290 & 1.372 & 7.16 & {}^{+1.5}_{-3.2} & \pm 3.   \\
 \hline 300 & 1.36 & 6.59 & {}^{+1.5}_{-3.2} & \pm 3.   \\
 \hline 310 & 1.348 & 6.08 & {}^{+1.5}_{-3.3} & \pm 3.   \\
 \hline 320 & 1.337 & 5.62 & {}^{+1.5}_{-3.2} & \pm 3.   \\
 \hline 330 & 1.326 & 5.2 & {}^{+1.5}_{-3.3} & \pm 3.   \\
 \hline 340 & 1.316 & 4.82 & {}^{+1.5}_{-3.3} & \pm 3.   \\
 \hline 350 & 1.306 & 4.48 & {}^{+1.5}_{-3.3} & \pm 3.   \\
 \hline 360 & 1.297 & 4.16 & {}^{+1.5}_{-3.3} & \pm 3.   \\
 \hline 370 & 1.287 & 3.88 & {}^{+1.6}_{-3.3} & \pm 3.   \\
 \hline 380 & 1.279 & 3.62 & {}^{+1.6}_{-3.3} & \pm 3.   \\
 \hline 390 & 1.27 & 3.38 & {}^{+1.6}_{-3.3} & \pm 3.1   \\
 \hline 400 & 1.262 & 3.16 & {}^{+1.6}_{-3.3} & \pm 3.1   \\
 \hline 410 & 1.254 & 2.96 & {}^{+1.6}_{-3.4} & \pm 3.1   \\
 \hline 420 & 1.246 & 2.77 & {}^{+1.6}_{-3.4} & \pm 3.1   \\
 \hline 430 & 1.239 & 2.6 & {}^{+1.6}_{-3.4} & \pm 3.1   \\
 \hline 440 & 1.232 & 2.44 & {}^{+1.7}_{-3.4} & \pm 3.1   \\
 \hline 450 & 1.225 & 2.3 & {}^{+1.6}_{-3.4} & \pm 3.1   \\
 \hline 460 & 1.218 & 2.16 & {}^{+1.7}_{-3.4} & \pm 3.2   \\
 \hline 470 & 1.211 & 2.03 & {}^{+1.7}_{-3.4} & \pm 3.2   \\
 \hline 480 & 1.205 & 1.92 & {}^{+1.7}_{-3.4} & \pm 3.2   \\
 \hline 490 & 1.199 & 1.81 & {}^{+1.7}_{-3.4} & \pm 3.2   \\
 \hline 500 & 1.193 & 1.71 & {}^{+1.7}_{-3.5} & \pm 3.3   \\
\hline
\end{array}\nonumber
\end{equation}
\caption{\label{tab:xsvalsmid} Production cross section for a scalar particle with a mass in the range 150~GeV to 500~GeV. }
\end{table}

\begin{table}[!h]
\begin{equation}
\begin{array}{|c|c|c|c|c|c|c|}
\hline
m_S \text{ [GeV]}&\hspace{0.2cm} C^2(\mu_0^2) \times 10^{-5}& \hspace{0.2cm} \sigma\text{ [fb]} \hspace{0.2cm}  & \hspace{0.2cm} \delta(\text{theory})\,[\%] \hspace{0.2cm}&  \hspace{0.2cm} \delta(\alpha_S+\text{PDF})\,[\%] \\
 \hline 500 & 1.193 & 1708.9 & {}^{+1.7}_{-3.5} & \pm 3.3   \\
 \hline 550 & 1.165 & 1297.2 & {}^{+1.8}_{-3.5} & \pm 3.4   \\
 \hline 600 & 1.141 & 1001. & {}^{+1.8}_{-3.6} & \pm 3.5   \\
 \hline 650 & 1.119 & 783.4 & {}^{+1.9}_{-3.6} & \pm 3.7   \\
 \hline 700 & 1.099 & 620.6 & {}^{+1.9}_{-3.7} & \pm 3.8   \\
 \hline 750 & 1.081 & 496.9 & {}^{+2.}_{-3.7} & \pm 4.   \\
 \hline 800 & 1.065 & 401.5 & {}^{+2.}_{-3.8} & \pm 4.2   \\
 \hline 850 & 1.05 & 327.1 & {}^{+2.1}_{-3.8} & \pm 4.4   \\
 \hline 900 & 1.036 & 268.5 & {}^{+2.1}_{-3.8} & \pm 4.6   \\
 \hline 950 & 1.023 & 221.9 & {}^{+2.2}_{-3.9} & \pm 4.8   \\
 \hline 1000 & 1.011 & 184.5 & {}^{+2.2}_{-4.} & \pm 5.   \\
 \hline 1050 & 1. & 154.2 & {}^{+2.2}_{-4.} & \pm 5.2   \\
 \hline 1100 & 0.989 & 129.5 & {}^{+2.3}_{-4.} & \pm 5.4   \\
 \hline 1150 & 0.979 & 109.3 & {}^{+2.3}_{-4.1} & \pm 5.6   \\
 \hline 1200 & 0.97 & 92.6 & {}^{+2.3}_{-4.1} & \pm 5.9   \\
 \hline 1250 & 0.961 & 78.8 & {}^{+2.3}_{-4.1} & \pm 6.1   \\
 \hline 1300 & 0.953 & 67.3 & {}^{+2.4}_{-4.2} & \pm 6.3   \\
 \hline 1350 & 0.945 & 57.6 & {}^{+2.3}_{-4.2} & \pm 6.6   \\
 \hline 1400 & 0.937 & 49.5 & {}^{+2.4}_{-4.2} & \pm 6.8   \\
 \hline 1450 & 0.93 & 42.7 & {}^{+2.4}_{-4.2} & \pm 7.   \\
 \hline 1500 & 0.923 & 36.9 & {}^{+2.4}_{-4.3} & \pm 7.3   \\
 \hline 1550 & 0.917 & 31.9 & {}^{+2.5}_{-4.3} & \pm 7.6   \\
 \hline 1600 & 0.91 & 27.7 & {}^{+2.5}_{-4.3} & \pm 7.8   \\
 \hline 1650 & 0.904 & 24.1 & {}^{+2.5}_{-4.4} & \pm 8.1   \\
 \hline 1700 & 0.898 & 21. & {}^{+2.5}_{-4.5} & \pm 8.5   \\
\hline
\end{array}\nonumber
\end{equation}
\caption{\label{tab:xsvalshigh} Production cross section for a scalar particle with a mass in the range 500~GeV to 1700~GeV. }
\end{table}

 \begin{table}[!h]
\begin{equation}
\begin{array}{|c|c|c|c|c|c|c|}
\hline
m_S \text{ [GeV]}&\hspace{0.2cm} C^2(\mu_0^2) \times 10^{-5}& \hspace{0.2cm} \sigma\text{ [fb]} \hspace{0.2cm}  & \hspace{0.2cm} \delta(\text{theory})\,[\%] \hspace{0.2cm}&  \hspace{0.2cm} \delta(\alpha_S+\text{PDF})\,[\%]  \\
\hline 1750 & 0.893 & 18.38 & {}^{+3.2}_{-4.4} & \pm 8.7   \\
 \hline 1800 & 0.887 & 16.09 & {}^{+2.5}_{-4.4} & \pm 8.9   \\
 \hline 1850 & 0.882 & 14.11 & {}^{+2.6}_{-4.4} & \pm 9.1   \\
 \hline 1900 & 0.877 & 12.39 & {}^{+2.5}_{-4.4} & \pm 9.4   \\
 \hline 1950 & 0.872 & 10.9 & {}^{+2.6}_{-4.5} & \pm 9.7   \\
 \hline 2000 & 0.868 & 9.6 & {}^{+2.6}_{-4.5} & \pm 9.7   \\
 \hline 2050 & 0.863 & 8.47 & {}^{+2.5}_{-4.5} & \pm 10.   \\
 \hline 2100 & 0.859 & 7.48 & {}^{+2.6}_{-4.5} & \pm 10.5   \\
 \hline 2150 & 0.855 & 6.62 & {}^{+2.6}_{-4.5} & \pm 10.8   \\
 \hline 2200 & 0.85 & 5.86 & {}^{+2.7}_{-4.6} & \pm 11.2   \\
 \hline 2250 & 0.846 & 5.19 & {}^{+2.6}_{-4.6} & \pm 11.8   \\
 \hline 2300 & 0.843 & 4.61 & {}^{+2.6}_{-4.5} & \pm 12.   \\
 \hline 2350 & 0.839 & 4.09 & {}^{+2.6}_{-4.5} & \pm 12.1   \\
 \hline 2400 & 0.835 & 3.64 & {}^{+2.6}_{-4.5} & \pm 12.4   \\
 \hline 2450 & 0.832 & 3.24 & {}^{+2.5}_{-4.5} & \pm 12.7   \\
 \hline 2500 & 0.828 & 2.89 & {}^{+2.6}_{-4.5} & \pm 13.1   \\
 \hline 2550 & 0.825 & 2.57 & {}^{+2.5}_{-4.5} & \pm 13.4   \\
 \hline 2600 & 0.821 & 2.3 & {}^{+2.5}_{-4.5} & \pm 13.7   \\
 \hline 2650 & 0.818 & 2.05 & {}^{+2.6}_{-4.6} & \pm 14.1   \\
 \hline 2700 & 0.815 & 1.83 & {}^{+2.7}_{-4.7} & \pm 14.5   \\
 \hline 2750 & 0.812 & 1.64 & {}^{+2.8}_{-4.8} & \pm 14.8   \\
 \hline 2800 & 0.809 & 1.46 & {}^{+2.9}_{-4.9} & \pm 15.2   \\
 \hline 2850 & 0.806 & 1.31 & {}^{+3.}_{-5.} & \pm 15.6   \\
 \hline 2900 & 0.803 & 1.17 & {}^{+3.1}_{-5.2} & \pm 15.9   \\
 \hline 2950 & 0.8 & 1.05 & {}^{+3.2}_{-5.3} & \pm 16.3   \\
 \hline 3000 & 0.798 & 0.94 & {}^{+3.4}_{-5.5} & \pm 16.7   \\
\hline
\end{array}\nonumber
\end{equation}
\caption{\label{tab:xsvalshigh2} Production cross section for a scalar particle with a mass in the range 1750~GeV to 3000~GeV. }
\end{table}

\begin{table}[!h]
\begin{equation}
\begin{array}{|c|c|c|c|c|c|c|}
\hline
m_S \text{ [GeV]}&\hspace{0.2cm} C^2(\mu_0^2) \times 10^{-5}& \hspace{0.2cm} \sigma\text{ [fb]} \hspace{0.2cm}  & \hspace{0.2cm} \delta(\text{theory})\,[\%] \hspace{0.2cm}&  \hspace{0.2cm} \delta(\alpha_S+\text{PDF})\,[\%]   \\
% \hline 720 & 1.092 & 567.1 & {}^{+2.}_{-3.7} & \pm 3.9   \\
% \hline 721 & 1.091 & 564.6 & {}^{+2.}_{-3.7} & \pm 3.9   \\
% \hline 722 & 1.091 & 562.1 & {}^{+2.}_{-3.7} & \pm 3.9   \\
% \hline 723 & 1.091 & 559.6 & {}^{+2.}_{-3.7} & \pm 3.9   \\
% \hline 724 & 1.09 & 557.1 & {}^{+2.}_{-3.7} & \pm 3.9   \\
% \hline 725 & 1.09 & 554.6 & {}^{+2.}_{-3.7} & \pm 3.9   \\
% \hline 726 & 1.089 & 552.1 & {}^{+2.}_{-3.7} & \pm 3.9   \\
% \hline 727 & 1.089 & 549.7 & {}^{+2.}_{-3.7} & \pm 3.9   \\
% \hline 728 & 1.089 & 547.3 & {}^{+2.}_{-3.7} & \pm 3.9   \\
% \hline 729 & 1.088 & 544.8 & {}^{+2.}_{-3.7} & \pm 3.9   \\
 \hline 730 & 1.088 & 542.4 & {}^{+2.}_{-3.7} & \pm 3.9   \\
 \hline 731 & 1.088 & 540. & {}^{+2.}_{-3.7} & \pm 3.9   \\
 \hline 732 & 1.087 & 537.7 & {}^{+2.}_{-3.7} & \pm 3.9   \\
 \hline 733 & 1.087 & 535.3 & {}^{+2.}_{-3.7} & \pm 4.   \\
 \hline 734 & 1.087 & 532.9 & {}^{+2.}_{-3.7} & \pm 4.   \\
 \hline 735 & 1.086 & 530.6 & {}^{+2.}_{-3.7} & \pm 4.   \\
 \hline 736 & 1.086 & 528.3 & {}^{+2.}_{-3.7} & \pm 4.   \\
 \hline 737 & 1.086 & 525.9 & {}^{+2.}_{-3.7} & \pm 4.   \\
 \hline 738 & 1.085 & 523.6 & {}^{+2.}_{-3.7} & \pm 4.   \\
 \hline 739 & 1.085 & 521.3 & {}^{+2.}_{-3.7} & \pm 4.   \\
 \hline 740 & 1.085 & 519. & {}^{+2.}_{-3.7} & \pm 4.   \\
 \hline 741 & 1.084 & 516.8 & {}^{+2.}_{-3.7} & \pm 4.   \\
 \hline 742 & 1.084 & 514.5 & {}^{+2.}_{-3.7} & \pm 4.   \\
 \hline 743 & 1.083 & 512.3 & {}^{+2.}_{-3.7} & \pm 4.   \\
 \hline 744 & 1.083 & 510. & {}^{+2.}_{-3.7} & \pm 4.   \\
 \hline 745 & 1.083 & 507.8 & {}^{+2.}_{-3.7} & \pm 4.   \\
 \hline 746 & 1.082 & 505.6 & {}^{+2.}_{-3.7} & \pm 4.   \\
 \hline 747 & 1.082 & 503.4 & {}^{+2.}_{-3.7} & \pm 4.   \\
 \hline 748 & 1.082 & 501.2 & {}^{+2.}_{-3.7} & \pm 4.   \\
 \hline 749 & 1.081 & 499. & {}^{+2.}_{-3.7} & \pm 4.   \\
 \hline 750 & 1.081 & 496.9 & {}^{+2.}_{-3.7} & \pm 4.   \\
 \hline 751 & 1.081 & 494.7 & {}^{+2.}_{-3.7} & \pm 4.   \\
 \hline 752 & 1.08 & 492.6 & {}^{+2.}_{-3.7} & \pm 4.   \\
 \hline 753 & 1.08 & 490.4 & {}^{+2.}_{-3.7} & \pm 4.   \\
 \hline 754 & 1.08 & 488.3 & {}^{+2.}_{-3.7} & \pm 4.   \\
 \hline 755 & 1.079 & 486.2 & {}^{+2.}_{-3.7} & \pm 4.   \\
 \hline 756 & 1.079 & 484.1 & {}^{+2.}_{-3.7} & \pm 4.   \\
 \hline 757 & 1.079 & 482. & {}^{+2.}_{-3.7} & \pm 4.   \\
 \hline 758 & 1.078 & 479.9 & {}^{+2.}_{-3.7} & \pm 4.   \\
 \hline 759 & 1.078 & 477.8 & {}^{+2.}_{-3.7} & \pm 4.   \\
 \hline 760 & 1.078 & 475.8 & {}^{+2.}_{-3.7} & \pm 4.   \\
 \hline 761 & 1.077 & 473.7 & {}^{+2.}_{-3.7} & \pm 4.   \\
 \hline 762 & 1.077 & 471.7 & {}^{+2.}_{-3.7} & \pm 4.   \\
 \hline 763 & 1.077 & 469.7 & {}^{+2.}_{-3.7} & \pm 4.   \\
 \hline 764 & 1.076 & 467.7 & {}^{+2.}_{-3.7} & \pm 4.   \\
 \hline 765 & 1.076 & 465.7 & {}^{+2.}_{-3.7} & \pm 4.   \\
 \hline 766 & 1.076 & 463.7 & {}^{+2.}_{-3.7} & \pm 4.   \\
 \hline 767 & 1.075 & 461.7 & {}^{+2.}_{-3.7} & \pm 4.   \\
 \hline 768 & 1.075 & 459.7 & {}^{+2.}_{-3.7} & \pm 4.   \\
 \hline 769 & 1.075 & 457.7 & {}^{+2.}_{-3.7} & \pm 4.   \\
 \hline 770 & 1.074 & 455.8 & {}^{+2.}_{-3.7} & \pm 4.   \\
% \hline 771 & 1.074 & 453.8 & {}^{+2.}_{-3.7} & \pm 4.   \\
% \hline 772 & 1.074 & 451.9 & {}^{+1.9}_{-3.7} & \pm 4.   \\
% \hline 773 & 1.073 & 450. & {}^{+1.9}_{-3.7} & \pm 3.9   \\
% \hline 774 & 1.073 & 448.1 & {}^{+1.9}_{-3.7} & \pm 4.   \\
% \hline 775 & 1.073 & 446.2 & {}^{+1.9}_{-3.7} & \pm 4.   \\
% \hline 776 & 1.072 & 444.3 & {}^{+1.9}_{-3.7} & \pm 4.   \\
% \hline 777 & 1.072 & 442.4 & {}^{+1.9}_{-3.7} & \pm 4.   \\
% \hline 778 & 1.072 & 440.5 & {}^{+1.9}_{-3.7} & \pm 4.   \\
% \hline 779 & 1.071 & 438.6 & {}^{+1.9}_{-3.7} & \pm 4.   \\
% \hline 780 & 1.071 & 436.8 & {}^{+1.9}_{-3.7} & \pm 4.1   \\
\hline
\end{array}\nonumber
\end{equation}
\caption{\label{tab:xsvals750} Production cross section for a scalar particle with a mass in the range 730~GeV to 770~GeV. }
\end{table}

 \begin{table}[!h]
 \small
\begin{equation}
\begin{array}{|c|c|c|c||c|c|c||c|c|c|}
\hline
m_S \text{ [GeV]}
& \sigma^{NLO}_S[1,1] [\textrm{pb}] & \delta_{\text{th}}\,[\%]  &  \delta_{\alpha_S}^{\text{PDF}}\,[\%]
&  \sigma^{NLO}_S[1,0][\textrm{pb}]  &\delta_{\text{th}}\,[\%]   &  \delta_{\alpha_S}^{\text{PDF}}\,[\%]
&  \sigma^{NLO}_S[0,1][\textrm{pb}] & \delta_{\text{th}}\,[\%]    & \delta_{\alpha_S}^{\text{PDF}}\,[\%]
\\ \hline

 50 & 687.1 & 34.22 & 4.24 & 171.4 & 34.22 & 4.25 & 172.3 & 34.24 & 4.25 \\\hline
 55 & 593.9 & 32.97 & 4.12 & 148.1 & 32.96 & 4.12 & 149. & 32.98 & 4.12 \\\hline
 60 & 518.3 & 31.93 & 4.02 & 129. & 31.92 & 4.02 & 130.2 & 31.94 & 4.02 \\\hline
 65 & 455.9 & 31.03 & 3.92 & 113.4 & 31.02 & 3.93 & 114.6 & 31.04 & 3.93 \\\hline
 70 & 404. & 30.23 & 3.84 & 100.4 & 30.22 & 3.84 & 101.7 & 30.25 & 3.85 \\\hline
 75 & 360.2 & 29.55 & 3.77 & 89.42 & 29.54 & 3.76 & 90.8 & 29.57 & 3.77 \\\hline
 80 & 323.1 & 28.91 & 3.71 & 80.07 & 28.9 & 3.71 & 81.56 & 28.93 & 3.71 \\\hline
 85 & 291.3 & 28.35 & 3.65 & 72.06 & 28.34 & 3.65 & 73.63 & 28.37 & 3.65 \\\hline
 90 & 263.8 & 27.84 & 3.6 & 65.15 & 27.83 & 3.6 & 66.8 & 27.86 & 3.6 \\\hline
 95 & 239.9 & 27.38 & 3.55 & 59.15 & 27.37 & 3.55 & 60.85 & 27.4 & 3.55 \\\hline
 100 & 219. & 26.96 & 3.51 & 53.9 & 26.96 & 3.51 & 55.66 & 26.98 & 3.51 \\\hline
 105 & 200.6 & 26.58 & 3.47 & 49.3 & 26.57 & 3.47 & 51.09 & 26.59 & 3.47 \\\hline
 110 & 184.4 & 26.22 & 3.43 & 45.22 & 26.21 & 3.43 & 47.06 & 26.23 & 3.43 \\\hline
 115 & 170. & 25.88 & 3.4 & 41.59 & 25.87 & 3.4 & 43.49 & 25.89 & 3.4 \\\hline
 120 & 157.2 & 25.57 & 3.37 & 38.35 & 25.56 & 3.37 & 40.3 & 25.57 & 3.37 \\\hline
 125 & 145.7 & 25.28 & 3.34 & 35.45 & 25.27 & 3.34 & 37.45 & 25.29 & 3.34 \\\hline
 130 & 135.4 & 25.01 & 3.31 & 32.86 & 25.01 & 3.32 & 34.89 & 25.02 & 3.31 \\\hline
 135 & 126.1 & 24.76 & 3.29 & 30.51 & 24.75 & 3.29 & 32.59 & 24.76 & 3.29 \\\hline
 140 & 117.7 & 24.51 & 3.28 & 28.41 & 24.51 & 3.28 & 30.51 & 24.52 & 3.27 \\\hline
 145 & 110.1 & 24.29 & 3.25 & 26.49 & 24.29 & 3.25 & 28.62 & 24.29 & 3.25 \\\hline
 150 & 103.1 & 24.07 & 3.23 & 24.74 & 24.07 & 3.23 & 26.91 & 24.07 & 3.23 \\\hline
 160 & 91.08 & 23.7 & 3.2 & 21.69 & 23.7 & 3.21 & 23.92 & 23.71 & 3.2 \\\hline
 170 & 80.94 & 23.36 & 3.17 & 19.12 & 23.34 & 3.18 & 21.42 & 23.38 & 3.17 \\\hline
 180 & 72.39 & 23.06 & 3.15 & 16.96 & 23.03 & 3.15 & 19.32 & 23.09 & 3.15 \\\hline
 190 & 65.1 & 22.78 & 3.13 & 15.11 & 22.74 & 3.13 & 17.53 & 22.83 & 3.13 \\\hline
 200 & 58.86 & 22.54 & 3.12 & 13.52 & 22.48 & 3.12 & 16. & 22.6 & 3.11 \\\hline
 210 & 53.49 & 22.31 & 3.11 & 12.15 & 22.23 & 3.11 & 14.69 & 22.39 & 3.1 \\\hline
 220 & 48.83 & 22.11 & 3.1 & 10.95 & 22.01 & 3.11 & 13.57 & 22.21 & 3.1 \\\hline
 230 & 44.79 & 21.93 & 3.1 & 9.911 & 21.81 & 3.1 & 12.6 & 22.05 & 3.09 \\\hline
 240 & 41.27 & 21.77 & 3.1 & 8.996 & 21.61 & 3.1 & 11.77 & 21.92 & 3.09 \\\hline
 250 & 38.2 & 21.63 & 3.1 & 8.189 & 21.43 & 3.1 & 11.05 & 21.8 & 3.09 \\\hline
 260 & 35.51 & 21.51 & 3.1 & 7.475 & 21.27 & 3.11 & 10.44 & 21.71 & 3.09 \\\hline
 270 & 33.17 & 21.4 & 3.11 & 6.84 & 21.11 & 3.11 & 9.919 & 21.65 & 3.1 \\\hline
 280 & 31.13 & 21.32 & 3.11 & 6.274 & 20.97 & 3.12 & 9.49 & 21.61 & 3.11 \\\hline
 290 & 29.38 & 21.26 & 3.12 & 5.767 & 20.83 & 3.13 & 9.148 & 21.61 & 3.11 \\\hline
 300 & 27.89 & 21.23 & 3.13 & 5.312 & 20.7 & 3.14 & 8.895 & 21.64 & 3.12 \\\hline
 310 & 26.68 & 21.24 & 3.14 & 4.903 & 20.58 & 3.15 & 8.742 & 21.74 & 3.13 \\\hline
 320 & 25.74 & 21.3 & 3.15 & 4.533 & 20.46 & 3.17 & 8.704 & 21.89 & 3.14 \\\hline
 330 & 25.59 & 21.75 & 3.17 & 4.199 & 20.35 & 3.19 & 9.09 & 22.67 & 3.16 \\\hline
 340 & 25.87 & 22.23 & 3.19 & 3.895 & 20.25 & 3.21 & 9.742 & 23.45 & 3.19 \\\hline
 350 & 25.37 & 21.6 & 3.22 & 3.62 & 20.15 & 3.23 & 9.915 & 22.53 & 3.21 \\\hline
 360 & 24.37 & 20.96 & 3.24 & 3.368 & 20.06 & 3.25 & 9.775 & 21.64 & 3.23 \\\hline
 370 & 23.1 & 20.47 & 3.26 & 3.139 & 19.97 & 3.27 & 9.437 & 20.99 & 3.26 \\\hline
 380 & 21.67 & 20.06 & 3.28 & 2.929 & 19.88 & 3.3 & 8.98 & 20.47 & 3.28 \\\hline
 390 & 20.19 & 19.88 & 3.31 & 2.736 & 19.8 & 3.32 & 8.457 & 20.05 & 3.3 \\\hline

 \end{array}\nonumber
\end{equation}
\caption{\label{tab:xsterms} NLO contributions to the production cross section for a 
scalar particle, as they are defined in eq.~\eqref{eq:master_bsm}. The theory uncertainty 
is computed as in eq.~\eqref{eq:NLO_err_incl_sc}}
\end{table}

 \begin{table}[!h]
 \small
\begin{equation}
\begin{array}{|c|c|c|c||c|c|c||c|c|c|}
\hline
m_S \text{ [GeV]}
& \sigma^{NLO}_S[1,1] [\textrm{pb}] & \delta_{\text{th}}\,[\%]  &  \delta_{\alpha_S}^{\text{PDF}}\,[\%]
&  \sigma^{NLO}_S[1,0] [\textrm{pb}] &\delta_{\text{th}}\,[\%]   &  \delta_{\alpha_S}^{\text{PDF}}\,[\%]
&  \sigma^{NLO}_S[0,1] [\textrm{pb}]& \delta_{\text{th}}\,[\%]    & \delta_{\alpha_S}^{\text{PDF}}\,[\%]
\\ \hline

 400 & 18.71 & 20.03 & 3.33 & 2.559 & 19.73 & 3.35 & 7.905 & 19.76 & 3.33 \\\hline
 410 & 17.28 & 20.14 & 3.35 & 2.397 & 19.65 & 3.37 & 7.347 & 19.92 & 3.35 \\\hline
 420 & 15.91 & 20.24 & 3.38 & 2.247 & 19.58 & 3.4 & 6.8 & 20.06 & 3.37 \\\hline
 430 & 14.61 & 20.31 & 3.4 & 2.109 & 19.51 & 3.43 & 6.275 & 20.17 & 3.4 \\\hline
 440 & 13.41 & 20.38 & 3.43 & 1.981 & 19.45 & 3.45 & 5.776 & 20.27 & 3.42 \\\hline
 450 & 12.29 & 20.43 & 3.46 & 1.863 & 19.39 & 3.48 & 5.308 & 20.35 & 3.45 \\\hline
 460 & 11.26 & 20.48 & 3.48 & 1.754 & 19.33 & 3.51 & 4.872 & 20.42 & 3.48 \\\hline
 470 & 10.31 & 20.52 & 3.51 & 1.652 & 19.27 & 3.54 & 4.467 & 20.48 & 3.5 \\\hline
 480 & 9.442 & 20.56 & 3.54 & 1.558 & 19.22 & 3.57 & 4.094 & 20.54 & 3.53 \\\hline
 490 & 8.647 & 20.59 & 3.57 & 1.47 & 19.17 & 3.6 & 3.751 & 20.59 & 3.56 \\\hline
 500 & 7.921 & 20.61 & 3.6 & 1.388 & 19.11 & 3.63 & 3.436 & 20.64 & 3.59 \\\hline
 510 & 7.258 & 20.64 & 3.62 & 1.312 & 19.07 & 3.66 & 3.147 & 20.68 & 3.62 \\\hline
 520 & 6.654 & 20.66 & 3.65 & 1.241 & 19.02 & 3.69 & 2.883 & 20.72 & 3.65 \\\hline
 530 & 6.104 & 20.68 & 3.68 & 1.175 & 18.98 & 3.72 & 2.642 & 20.75 & 3.68 \\\hline
 540 & 5.602 & 20.69 & 3.72 & 1.113 & 18.93 & 3.75 & 2.422 & 20.79 & 3.71 \\\hline
 550 & 5.145 & 20.7 & 3.75 & 1.055 & 18.89 & 3.78 & 2.221 & 20.82 & 3.74 \\\hline
 560 & 4.729 & 20.71 & 3.78 & 1. & 18.85 & 3.82 & 2.037 & 20.85 & 3.77 \\\hline
 570 & 4.349 & 20.73 & 3.81 & 0.9494 & 18.82 & 3.85 & 1.87 & 20.88 & 3.8 \\\hline
 580 & 4.003 & 20.73 & 3.84 & 0.9016 & 18.78 & 3.88 & 1.717 & 20.9 & 3.83 \\\hline
 590 & 3.687 & 20.74 & 3.87 & 0.8567 & 18.74 & 3.91 & 1.578 & 20.93 & 3.86 \\\hline
 600 & 3.399 & 20.74 & 3.9 & 0.8145 & 18.7 & 3.95 & 1.451 & 20.95 & 3.9 \\\hline
 610 & 3.136 & 20.75 & 3.94 & 0.7748 & 18.67 & 3.98 & 1.335 & 20.98 & 3.93 \\\hline
 620 & 2.896 & 20.75 & 3.97 & 0.7374 & 18.64 & 4.01 & 1.228 & 21. & 3.96 \\\hline
 630 & 2.677 & 20.74 & 4. & 0.7022 & 18.61 & 4.04 & 1.131 & 21.02 & 3.99 \\\hline
 640 & 2.476 & 20.75 & 4.04 & 0.669 & 18.58 & 4.08 & 1.043 & 21.04 & 4.03 \\\hline
 650 & 2.292 & 20.74 & 4.07 & 0.6377 & 18.54 & 4.11 & 0.9616 & 21.05 & 4.06 \\\hline
 660 & 2.123 & 20.74 & 4.1 & 0.6082 & 18.52 & 4.14 & 0.8873 & 21.08 & 4.09 \\\hline
 670 & 1.969 & 20.74 & 4.14 & 0.5803 & 18.49 & 4.18 & 0.8193 & 21.1 & 4.13 \\\hline
 680 & 1.827 & 20.73 & 4.17 & 0.5539 & 18.47 & 4.21 & 0.757 & 21.12 & 4.16 \\\hline
 690 & 1.697 & 20.73 & 4.2 & 0.5289 & 18.44 & 4.25 & 0.6998 & 21.14 & 4.19 \\\hline
 700 & 1.577 & 20.72 & 4.24 & 0.5053 & 18.42 & 4.28 & 0.6474 & 21.16 & 4.23 \\\hline
 710 & 1.467 & 20.72 & 4.27 & 0.483 & 18.39 & 4.31 & 0.5993 & 21.18 & 4.26 \\\hline
 720 & 1.366 & 20.71 & 4.31 & 0.4618 & 18.37 & 4.35 & 0.5551 & 21.2 & 4.3 \\\hline
 730 & 1.273 & 20.7 & 4.34 & 0.4417 & 18.35 & 4.38 & 0.5144 & 21.21 & 4.33 \\\hline
 740 & 1.187 & 20.69 & 4.37 & 0.4227 & 18.32 & 4.42 & 0.477 & 21.23 & 4.36 \\\hline
 750 & 1.108 & 20.68 & 4.41 & 0.4046 & 18.3 & 4.45 & 0.4427 & 21.25 & 4.4 \\\hline
 760 & 1.035 & 20.67 & 4.44 & 0.3875 & 18.28 & 4.49 & 0.411 & 21.27 & 4.43 \\\hline
 770 & 0.967 & 20.66 & 4.48 & 0.3712 & 18.26 & 4.52 & 0.3818 & 21.29 & 4.47 \\\hline
 780 & 0.9045 & 20.65 & 4.51 & 0.3557 & 18.24 & 4.55 & 0.3549 & 21.31 & 4.5 \\\hline
 790 & 0.8467 & 20.63 & 4.55 & 0.341 & 18.22 & 4.59 & 0.3301 & 21.32 & 4.54 \\\hline
 800 & 0.7932 & 20.62 & 4.58 & 0.327 & 18.21 & 4.62 & 0.3071 & 21.34 & 4.57 \\\hline
 810 & 0.7435 & 20.61 & 4.62 & 0.3137 & 18.19 & 4.66 & 0.286 & 21.36 & 4.61 \\\hline
 820 & 0.6976 & 20.6 & 4.65 & 0.301 & 18.17 & 4.69 & 0.2664 & 21.38 & 4.64 \\\hline
 830 & 0.6549 & 20.58 & 4.69 & 0.2889 & 18.16 & 4.74 & 0.2483 & 21.4 & 4.68 \\\hline
 840 & 0.6152 & 20.57 & 4.72 & 0.2774 & 18.14 & 4.76 & 0.2316 & 21.42 & 4.71 \\\hline

\end{array}\nonumber
\end{equation}
\caption{\label{tab:xsterms2} NLO contributions to the production cross section 
for a scalar particle, as they are defined in eq.~\eqref{eq:master_bsm}. The theory 
uncertainty is computed as in eq.~\eqref{eq:NLO_err_incl_sc}. }
\end{table}

 \begin{table}[!h]
 \small
\begin{equation}
\begin{array}{|c|c|c|c||c|c|c||c|c|c|}
\hline
m_S \text{ [GeV]}
& \sigma^{NLO}_S[1,1] [\textrm{fb}] & \delta_{\text{th}}\,[\%]  &  \delta_{\alpha_S}^{\text{PDF}}\,[\%]
&  \sigma^{NLO}_S[1,0] [\textrm{fb}] &\delta_{\text{th}}\,[\%]   &  \delta_{\alpha_S}^{\text{PDF}}\,[\%]
&  \sigma^{NLO}_S[0,1] [\textrm{fb}]& \delta_{\text{th}}\,[\%]    & \delta_{\alpha_S}^{\text{PDF}}\,[\%]
\\ \hline

 850 & 578.4 & 20.55 & 4.76 & 266.4 & 18.12 & 4.8 & 216.1 & 21.43 & 4.75 \\\hline
 860 & 544.1 & 20.53 & 4.79 & 256. & 18.11 & 4.83 & 201.7 & 21.45 & 4.79 \\\hline
 870 & 512.2 & 20.51 & 4.83 & 246. & 18.09 & 4.87 & 188.4 & 21.47 & 4.82 \\\hline
 880 & 482.5 & 20.49 & 4.87 & 236.4 & 18.08 & 4.91 & 176.1 & 21.48 & 4.86 \\\hline
 890 & 454.8 & 20.47 & 4.9 & 227.3 & 18.06 & 4.94 & 164.6 & 21.5 & 4.89 \\\hline
 900 & 429. & 20.45 & 4.94 & 218.7 & 18.05 & 4.98 & 154. & 21.52 & 4.93 \\\hline
 910 & 404.9 & 20.44 & 4.97 & 210.4 & 18.04 & 5.01 & 144.1 & 21.54 & 4.97 \\\hline
 920 & 382.4 & 20.41 & 5.01 & 202.4 & 18.02 & 5.05 & 134.9 & 21.55 & 5. \\\hline
 930 & 361.4 & 20.4 & 5.04 & 194.8 & 18.01 & 5.08 & 126.4 & 21.57 & 5.04 \\\hline
 940 & 341.7 & 20.38 & 5.08 & 187.6 & 18. & 5.12 & 118.4 & 21.59 & 5.07 \\\hline
 950 & 323.3 & 20.36 & 5.14 & 180.6 & 17.99 & 5.15 & 111. & 21.61 & 5.11 \\\hline
 960 & 306.1 & 20.33 & 5.19 & 174. & 17.98 & 5.19 & 104.1 & 21.63 & 5.15 \\\hline
 970 & 290. & 20.31 & 5.22 & 167.6 & 17.97 & 5.23 & 97.73 & 21.65 & 5.18 \\\hline
 980 & 274.8 & 20.29 & 5.25 & 161.5 & 17.96 & 5.26 & 91.74 & 21.66 & 5.22 \\\hline
 990 & 260.6 & 20.27 & 5.26 & 155.7 & 17.95 & 5.3 & 86.16 & 21.68 & 5.26 \\\hline
 1000 & 247.3 & 20.25 & 5.3 & 150.1 & 17.94 & 5.33 & 80.95 & 21.7 & 5.29 \\\hline
 1010 & 234.8 & 20.23 & 5.33 & 144.8 & 17.93 & 5.37 & 76.09 & 21.72 & 5.33 \\\hline
 1020 & 223. & 20.2 & 5.37 & 139.6 & 17.92 & 5.41 & 71.55 & 21.74 & 5.37 \\\hline
 1030 & 212. & 20.18 & 5.41 & 134.7 & 17.91 & 5.44 & 67.3 & 21.76 & 5.4 \\\hline
 1040 & 201.6 & 20.15 & 5.44 & 130. & 17.91 & 5.48 & 63.34 & 21.78 & 5.44 \\\hline
 1050 & 191.8 & 20.13 & 5.48 & 125.4 & 17.9 & 5.52 & 59.63 & 21.8 & 5.48 \\\hline
 1060 & 182.5 & 20.11 & 5.52 & 121.1 & 17.89 & 5.55 & 56.15 & 21.82 & 5.51 \\\hline
 1070 & 173.8 & 20.08 & 5.55 & 116.9 & 17.88 & 5.59 & 52.9 & 21.83 & 5.55 \\\hline
 1080 & 165.6 & 20.06 & 5.59 & 112.9 & 17.87 & 5.63 & 49.86 & 21.85 & 5.59 \\\hline
 1090 & 157.9 & 20.03 & 5.63 & 109. & 17.87 & 5.66 & 47. & 21.87 & 5.63 \\\hline
 1100 & 150.6 & 20.01 & 5.66 & 105.3 & 17.86 & 5.7 & 44.33 & 21.89 & 5.66 \\\hline
 1110 & 143.7 & 19.98 & 5.7 & 101.8 & 17.85 & 5.74 & 41.82 & 21.91 & 5.7 \\\hline
 1120 & 137.2 & 19.96 & 5.74 & 98.33 & 17.85 & 5.77 & 39.47 & 21.93 & 5.74 \\\hline
 1130 & 131. & 19.93 & 5.77 & 95.04 & 17.84 & 5.81 & 37.27 & 21.94 & 5.78 \\\hline
 1140 & 125.2 & 19.9 & 5.81 & 91.87 & 17.83 & 5.85 & 35.19 & 21.96 & 5.81 \\\hline
 1150 & 119.6 & 19.88 & 5.85 & 88.83 & 17.83 & 5.88 & 33.25 & 21.98 & 5.85 \\\hline
 1160 & 114.4 & 19.85 & 5.88 & 85.9 & 17.82 & 5.92 & 31.42 & 22. & 5.89 \\\hline
 1170 & 109.4 & 19.83 & 5.92 & 83.08 & 17.82 & 5.96 & 29.7 & 22.02 & 5.93 \\\hline
 1180 & 104.7 & 19.8 & 5.96 & 80.36 & 17.81 & 6. & 28.09 & 22.04 & 5.97 \\\hline
 1190 & 100.3 & 19.78 & 6. & 77.75 & 17.81 & 6.03 & 26.57 & 22.06 & 6. \\\hline
 1200 & 96.05 & 19.75 & 6.03 & 75.23 & 17.8 & 6.07 & 25.14 & 22.07 & 6.04 \\\hline
 1210 & 92.03 & 19.73 & 6.07 & 72.81 & 17.8 & 6.11 & 23.8 & 22.1 & 6.08 \\\hline
 1220 & 88.22 & 19.7 & 6.11 & 70.47 & 17.8 & 6.15 & 22.53 & 22.12 & 6.12 \\\hline
 1230 & 84.59 & 19.68 & 6.15 & 68.22 & 17.79 & 6.18 & 21.34 & 22.13 & 6.16 \\\hline
 1240 & 81.15 & 19.65 & 6.19 & 66.05 & 17.79 & 6.22 & 20.21 & 22.15 & 6.2 \\\hline
 1250 & 77.87 & 19.63 & 6.22 & 63.96 & 17.78 & 6.26 & 19.15 & 22.17 & 6.24 \\\hline
 1260 & 74.75 & 19.6 & 6.26 & 61.94 & 17.78 & 6.3 & 18.16 & 22.19 & 6.27 \\\hline
 1270 & 71.78 & 19.57 & 6.3 & 60. & 17.77 & 6.34 & 17.21 & 22.2 & 6.31 \\\hline
 1280 & 68.96 & 19.55 & 6.34 & 58.12 & 17.77 & 6.37 & 16.33 & 22.23 & 6.35 \\\hline
 1290 & 66.26 & 19.52 & 6.38 & 56.31 & 17.76 & 6.41 & 15.49 & 22.24 & 6.39 \\\hline

\end{array}\nonumber
\end{equation}
\caption{\label{tab:xsterms3} NLO contributions to the production cross section for a scalar particle, as they are defined in eq.~\eqref{eq:master_bsm}. The theory uncertainty is computed as in eq.~\eqref{eq:NLO_err_incl_sc}. }
\end{table}

\begin{table}[!h]
\small
\begin{equation}
\begin{array}{|c|c|c|c||c|c|c||c|c|c|}
\hline
m_S \text{ [GeV]}
& \sigma^{NLO}_S[1,1] [\textrm{fb}] & \delta_{\text{th}}\,[\%]  &  \delta_{\alpha_S}^{\text{PDF}}\,[\%]
&  \sigma^{NLO}_S[1,0][\textrm{fb}]  &\delta_{\text{th}}\,[\%]   &  \delta_{\alpha_S}^{\text{PDF}}\,[\%]
&  \sigma^{NLO}_S[0,1][\textrm{fb}] & \delta_{\text{th}}\,[\%]    & \delta_{\alpha_S}^{\text{PDF}}\,[\%]
\\ \hline

 1300 & 63.7 & 19.49 & 6.41 & 54.57 & 17.76 & 6.45 & 14.7 & 22.26 & 6.43 \\\hline
 1310 & 61.23 & 19.48 & 6.46 & 52.88 & 17.76 & 6.49 & 13.95 & 22.28 & 6.47 \\\hline
 1320 & 58.89 & 19.45 & 6.5 & 51.26 & 17.76 & 6.53 & 13.24 & 22.3 & 6.51 \\\hline
 1330 & 56.65 & 19.44 & 6.54 & 49.69 & 17.75 & 6.57 & 12.58 & 22.32 & 6.55 \\\hline
 1340 & 54.52 & 19.41 & 6.57 & 48.17 & 17.75 & 6.61 & 11.95 & 22.34 & 6.59 \\\hline
 1350 & 52.5 & 19.39 & 6.6 & 46.71 & 17.75 & 6.64 & 11.35 & 22.36 & 6.63 \\\hline
 1360 & 50.59 & 19.35 & 6.62 & 45.3 & 17.75 & 6.68 & 10.79 & 22.38 & 6.67 \\\hline
 1370 & 48.73 & 19.33 & 6.67 & 43.93 & 17.74 & 6.72 & 10.25 & 22.4 & 6.71 \\\hline
 1380 & 46.96 & 19.31 & 6.72 & 42.61 & 17.74 & 6.76 & 9.749 & 22.42 & 6.75 \\\hline
 1390 & 45.26 & 19.28 & 6.76 & 41.33 & 17.74 & 6.8 & 9.272 & 22.43 & 6.79 \\\hline
 1400 & 43.64 & 19.26 & 6.8 & 40.1 & 17.74 & 6.84 & 8.82 & 22.46 & 6.83 \\\hline
 1410 & 42.09 & 19.24 & 6.84 & 38.91 & 17.74 & 6.88 & 8.392 & 22.48 & 6.87 \\\hline
 1420 & 40.6 & 19.22 & 6.9 & 37.76 & 17.74 & 6.92 & 7.986 & 22.5 & 6.91 \\\hline
 1430 & 39.17 & 19.2 & 6.95 & 36.65 & 17.74 & 6.96 & 7.602 & 22.52 & 6.95 \\\hline
 1440 & 37.81 & 19.17 & 6.99 & 35.57 & 17.74 & 7. & 7.238 & 22.54 & 6.99 \\\hline
 1450 & 36.5 & 19.16 & 7.03 & 34.53 & 17.74 & 7.04 & 6.893 & 22.57 & 7.03 \\\hline
 1460 & 35.24 & 19.13 & 7.06 & 33.52 & 17.74 & 7.08 & 6.566 & 22.58 & 7.07 \\\hline
 1470 & 34.04 & 19.11 & 7.08 & 32.55 & 17.74 & 7.12 & 6.256 & 22.6 & 7.11 \\\hline
 1480 & 32.89 & 19.09 & 7.12 & 31.6 & 17.74 & 7.16 & 5.961 & 22.62 & 7.15 \\\hline
 1490 & 31.78 & 19.07 & 7.16 & 30.69 & 17.74 & 7.2 & 5.682 & 22.64 & 7.19 \\\hline
 1500 & 30.72 & 19.05 & 7.2 & 29.81 & 17.74 & 7.24 & 5.417 & 22.66 & 7.23 \\\hline
 1510 & 29.69 & 19.03 & 7.24 & 28.95 & 17.74 & 7.28 & 5.165 & 22.68 & 7.28 \\\hline
 1520 & 28.71 & 19.01 & 7.28 & 28.12 & 17.74 & 7.32 & 4.926 & 22.7 & 7.32 \\\hline
 1530 & 27.77 & 18.99 & 7.32 & 27.32 & 17.74 & 7.36 & 4.699 & 22.72 & 7.36 \\\hline
 1540 & 26.86 & 18.97 & 7.36 & 26.55 & 17.74 & 7.4 & 4.483 & 22.74 & 7.4 \\\hline
 1550 & 25.99 & 18.96 & 7.4 & 25.8 & 17.74 & 7.44 & 4.278 & 22.76 & 7.44 \\\hline
 1560 & 25.15 & 18.94 & 7.44 & 25.07 & 17.74 & 7.48 & 4.083 & 22.78 & 7.48 \\\hline
 1570 & 24.35 & 18.92 & 7.48 & 24.36 & 17.74 & 7.52 & 3.898 & 22.8 & 7.53 \\\hline
 1580 & 23.57 & 18.9 & 7.52 & 23.68 & 17.74 & 7.56 & 3.722 & 22.82 & 7.57 \\\hline
 1590 & 22.83 & 18.88 & 7.56 & 23.02 & 17.74 & 7.61 & 3.555 & 22.85 & 7.61 \\\hline
 1600 & 22.11 & 18.87 & 7.6 & 22.38 & 17.74 & 7.65 & 3.395 & 22.87 & 7.65 \\\hline
 1610 & 21.42 & 18.85 & 7.65 & 21.75 & 17.74 & 7.69 & 3.244 & 22.89 & 7.7 \\\hline
 1620 & 20.75 & 18.83 & 7.69 & 21.15 & 17.74 & 7.73 & 3.099 & 22.9 & 7.74 \\\hline
 1630 & 20.11 & 18.81 & 7.73 & 20.57 & 17.75 & 7.77 & 2.962 & 22.93 & 7.78 \\\hline
 1640 & 19.49 & 18.8 & 7.77 & 20. & 17.75 & 7.81 & 2.831 & 22.95 & 7.82 \\\hline
 1650 & 18.89 & 18.78 & 7.81 & 19.45 & 17.75 & 7.86 & 2.707 & 22.97 & 7.87 \\\hline
 1660 & 18.31 & 18.77 & 7.86 & 18.92 & 17.75 & 7.9 & 2.588 & 22.99 & 7.91 \\\hline
 1670 & 17.76 & 18.75 & 7.9 & 18.4 & 17.75 & 7.94 & 2.475 & 23.01 & 7.95 \\\hline
 1680 & 17.22 & 18.74 & 7.94 & 17.9 & 17.76 & 7.98 & 2.367 & 23.03 & 8. \\\hline
 1690 & 16.71 & 18.72 & 7.98 & 17.41 & 17.76 & 8.03 & 2.265 & 23.05 & 8.04 \\\hline
 1700 & 16.21 & 18.7 & 8.02 & 16.94 & 17.75 & 8.07 & 2.167 & 23.07 & 8.08 \\\hline
 1710 & 15.73 & 18.69 & 8.07 & 16.48 & 17.76 & 8.11 & 2.074 & 23.09 & 8.13 \\\hline
 1720 & 15.26 & 18.68 & 8.11 & 16.04 & 17.76 & 8.16 & 1.985 & 23.11 & 8.17 \\\hline
 1730 & 14.81 & 18.66 & 8.15 & 15.61 & 17.76 & 8.2 & 1.9 & 23.13 & 8.22 \\\hline
 1740 & 14.38 & 18.65 & 8.2 & 15.19 & 17.76 & 8.24 & 1.819 & 23.15 & 8.26 \\\hline

\end{array}\nonumber
\end{equation}
\caption{\label{tab:xsterms4} NLO contributions to the production cross section for a scalar particle, as they are defined in eq.~\eqref{eq:master_bsm}. The theory uncertainty is computed as in eq.~\eqref{eq:NLO_err_incl_sc}. }
\end{table}

\begin{table}[!h]
\small
\begin{equation}
\begin{array}{|c|c|c|c||c|c|c||c|c|c|}
\hline
m_S \text{ [GeV]}
& \sigma^{NLO}_S[1,1][\textrm{fb}] & \delta_{\text{th}}\,[\%]  &  \delta_{\alpha_S}^{\text{PDF}}\,[\%]
&  \sigma^{NLO}_S[1,0][\textrm{fb}]  &\delta_{\text{th}}\,[\%]   &  \delta_{\alpha_S}^{\text{PDF}}\,[\%]
&  \sigma^{NLO}_S[0,1][\textrm{fb}] & \delta_{\text{th}}\,[\%]    & \delta_{\alpha_S}^{\text{PDF}}\,[\%]
\\ \hline

 1750 & 13.96 & 18.64 & 8.24 & 14.79 & 17.77 & 8.29 & 1.742 & 23.18 & 8.31 \\\hline
 1760 & 13.56 & 18.63 & 8.28 & 14.39 & 17.77 & 8.33 & 1.668 & 23.2 & 8.35 \\\hline
 1770 & 13.17 & 18.61 & 8.33 & 14.01 & 17.77 & 8.37 & 1.598 & 23.22 & 8.39 \\\hline
 1780 & 12.79 & 18.6 & 8.37 & 13.64 & 17.78 & 8.42 & 1.531 & 23.24 & 8.44 \\\hline
 1790 & 12.42 & 18.59 & 8.41 & 13.28 & 17.78 & 8.46 & 1.467 & 23.26 & 8.48 \\\hline
 1800 & 12.07 & 18.57 & 8.46 & 12.93 & 17.78 & 8.5 & 1.405 & 23.28 & 8.53 \\\hline
 1810 & 11.73 & 18.56 & 8.5 & 12.59 & 17.78 & 8.55 & 1.347 & 23.3 & 8.57 \\\hline
 1820 & 11.4 & 18.55 & 8.54 & 12.26 & 17.79 & 8.59 & 1.291 & 23.32 & 8.62 \\\hline
 1830 & 11.08 & 18.54 & 8.59 & 11.94 & 17.79 & 8.64 & 1.238 & 23.34 & 8.66 \\\hline
 1840 & 10.77 & 18.53 & 8.63 & 11.62 & 17.79 & 8.68 & 1.187 & 23.36 & 8.71 \\\hline
 1850 & 10.47 & 18.52 & 8.68 & 11.32 & 17.8 & 8.73 & 1.138 & 23.39 & 8.76 \\\hline
 1860 & 10.18 & 18.51 & 8.72 & 11.03 & 17.8 & 8.77 & 1.092 & 23.41 & 8.8 \\\hline
 1870 & 9.894 & 18.5 & 8.77 & 10.74 & 17.8 & 8.82 & 1.047 & 23.43 & 8.85 \\\hline
 1880 & 9.621 & 18.49 & 8.81 & 10.46 & 17.8 & 8.86 & 1.004 & 23.45 & 8.89 \\\hline
 1890 & 9.358 & 18.48 & 8.86 & 10.19 & 17.81 & 8.91 & 0.9637 & 23.47 & 8.94 \\\hline
 1900 & 9.102 & 18.47 & 8.9 & 9.93 & 17.81 & 8.95 & 0.9248 & 23.49 & 8.99 \\\hline
 1910 & 8.854 & 18.46 & 8.95 & 9.675 & 17.82 & 9. & 0.8875 & 23.52 & 9.03 \\\hline
 1920 & 8.614 & 18.45 & 8.99 & 9.427 & 17.82 & 9.04 & 0.8519 & 23.54 & 9.08 \\\hline
 1930 & 8.382 & 18.44 & 9.04 & 9.186 & 17.82 & 9.09 & 0.8178 & 23.56 & 9.13 \\\hline
 1940 & 8.156 & 18.44 & 9.08 & 8.951 & 17.83 & 9.14 & 0.7851 & 23.58 & 9.17 \\\hline
 1950 & 7.937 & 18.43 & 9.13 & 8.723 & 17.83 & 9.18 & 0.7539 & 23.6 & 9.22 \\\hline
 1960 & 7.725 & 18.42 & 9.18 & 8.502 & 17.83 & 9.23 & 0.724 & 23.62 & 9.27 \\\hline
 1970 & 7.52 & 18.41 & 9.22 & 8.286 & 17.84 & 9.27 & 0.6953 & 23.64 & 9.32 \\\hline
 1980 & 7.32 & 18.4 & 9.27 & 8.076 & 17.84 & 9.32 & 0.6679 & 23.67 & 9.37 \\\hline
 1990 & 7.127 & 18.4 & 9.32 & 7.872 & 17.85 & 9.37 & 0.6416 & 23.69 & 9.41 \\\hline
 2000 & 6.939 & 18.39 & 9.36 & 7.674 & 17.85 & 9.42 & 0.6164 & 23.71 & 9.46 \\\hline
 2010 & 6.757 & 18.38 & 9.41 & 7.481 & 17.86 & 9.46 & 0.5923 & 23.73 & 9.51 \\\hline
 2020 & 6.58 & 18.38 & 9.46 & 7.293 & 17.86 & 9.51 & 0.5692 & 23.75 & 9.56 \\\hline
 2030 & 6.408 & 18.37 & 9.5 & 7.111 & 17.86 & 9.56 & 0.5471 & 23.77 & 9.61 \\\hline
 2040 & 6.241 & 18.36 & 9.55 & 6.933 & 17.87 & 9.6 & 0.5259 & 23.79 & 9.65 \\\hline
 2050 & 6.08 & 18.36 & 9.6 & 6.76 & 17.87 & 9.65 & 0.5056 & 23.82 & 9.7 \\\hline
 2060 & 5.923 & 18.35 & 9.65 & 6.592 & 17.88 & 9.7 & 0.4861 & 23.84 & 9.75 \\\hline
 2070 & 5.77 & 18.35 & 9.69 & 6.428 & 17.88 & 9.75 & 0.4674 & 23.86 & 9.8 \\\hline
 2080 & 5.622 & 18.34 & 9.74 & 6.269 & 17.89 & 9.8 & 0.4495 & 23.88 & 9.85 \\\hline
 2090 & 5.478 & 18.34 & 9.79 & 6.114 & 17.89 & 9.85 & 0.4323 & 23.91 & 9.9 \\\hline
 2100 & 5.338 & 18.33 & 9.84 & 5.963 & 17.9 & 9.89 & 0.4158 & 23.93 & 9.95 \\\hline
 2110 & 5.202 & 18.33 & 9.89 & 5.816 & 17.9 & 9.94 & 0.4 & 23.95 & 10. \\\hline
 2120 & 5.07 & 18.33 & 9.94 & 5.673 & 17.91 & 9.99 & 0.3848 & 23.98 & 10.05 \\\hline
 2130 & 4.942 & 18.32 & 9.98 & 5.533 & 17.92 & 10.04 & 0.3702 & 24. & 10.1 \\\hline
 2140 & 4.817 & 18.32 & 10.03 & 5.397 & 17.92 & 10.09 & 0.3562 & 24.02 & 10.15 \\\hline
 2150 & 4.696 & 18.31 & 10.08 & 5.265 & 17.93 & 10.14 & 0.3428 & 24.04 & 10.2 \\\hline
 2160 & 4.578 & 18.31 & 10.13 & 5.137 & 17.93 & 10.19 & 0.33 & 24.07 & 10.25 \\\hline
 2170 & 4.463 & 18.31 & 10.18 & 5.011 & 17.94 & 10.24 & 0.3176 & 24.09 & 10.3 \\\hline
 2180 & 4.352 & 18.31 & 10.23 & 4.889 & 17.94 & 10.29 & 0.3057 & 24.11 & 10.35 \\\hline
 2190 & 4.244 & 18.3 & 10.28 & 4.771 & 17.95 & 10.34 & 0.2944 & 24.14 & 10.4 \\\hline

\end{array}\nonumber
\end{equation}
\caption{\label{tab:xsterms5} NLO contributions to the production cross section for a scalar particle, as they are defined in eq.~\eqref{eq:master_bsm}. The theory uncertainty is computed as in eq.~\eqref{eq:NLO_err_incl_sc}. }
\end{table}

\begin{table}[!h]
\small
\begin{equation}
\begin{array}{|c|c|c|c||c|c|c||c|c|c|}
\hline
m_S \text{ [GeV]}
& \sigma^{NLO}_S[1,1] [\textrm{fb}] & \delta_{\text{th}}\,[\%]  &  \delta_{\alpha_S}^{\text{PDF}}\,[\%]
&  \sigma^{NLO}_S[1,0][\textrm{fb}]  &\delta_{\text{th}}\,[\%]   &  \delta_{\alpha_S}^{\text{PDF}}\,[\%]
&  \sigma^{NLO}_S[0,1][\textrm{fb}] & \delta_{\text{th}}\,[\%]    & \delta_{\alpha_S}^{\text{PDF}}\,[\%]
\\ \hline

 2200 & 4.138 & 18.3 & 10.33 & 4.655 & 17.96 & 10.39 & 0.2834 & 24.16 & 10.46 \\\hline
 2210 & 4.036 & 18.3 & 10.38 & 4.542 & 17.96 & 10.44 & 0.2729 & 24.18 & 10.51 \\\hline
 2220 & 3.936 & 18.29 & 10.43 & 4.432 & 17.97 & 10.49 & 0.2628 & 24.2 & 10.56 \\\hline
 2230 & 3.839 & 18.29 & 10.48 & 4.325 & 17.97 & 10.54 & 0.2531 & 24.23 & 10.61 \\\hline
 2240 & 3.744 & 18.29 & 10.53 & 4.221 & 17.98 & 10.59 & 0.2438 & 24.25 & 10.66 \\\hline
 2250 & 3.653 & 18.29 & 10.58 & 4.12 & 17.99 & 10.64 & 0.2349 & 24.27 & 10.72 \\\hline
 2260 & 3.563 & 18.29 & 10.63 & 4.021 & 17.99 & 10.69 & 0.2263 & 24.3 & 10.77 \\\hline
 2270 & 3.476 & 18.29 & 10.69 & 3.924 & 18. & 10.74 & 0.218 & 24.32 & 10.82 \\\hline
 2280 & 3.392 & 18.29 & 10.74 & 3.831 & 18.01 & 10.8 & 0.21 & 24.35 & 10.87 \\\hline
 2290 & 3.309 & 18.28 & 10.79 & 3.739 & 18.01 & 10.85 & 0.2024 & 24.37 & 10.93 \\\hline
 2300 & 3.229 & 18.28 & 10.84 & 3.65 & 18.02 & 10.9 & 0.1951 & 24.39 & 10.98 \\\hline
 2310 & 3.151 & 18.28 & 10.89 & 3.563 & 18.02 & 10.95 & 0.188 & 24.41 & 11.03 \\\hline
 2320 & 3.075 & 18.28 & 10.94 & 3.478 & 18.03 & 11. & 0.1812 & 24.44 & 11.09 \\\hline
 2330 & 3.001 & 18.28 & 11. & 3.396 & 18.04 & 11.06 & 0.1747 & 24.46 & 11.14 \\\hline
 2340 & 2.929 & 18.28 & 11.05 & 3.315 & 18.04 & 11.11 & 0.1684 & 24.48 & 11.19 \\\hline
 2350 & 2.858 & 18.28 & 11.1 & 3.237 & 18.05 & 11.16 & 0.1624 & 24.51 & 11.25 \\\hline
 2360 & 2.79 & 18.28 & 11.15 & 3.16 & 18.06 & 11.21 & 0.1566 & 24.53 & 11.3 \\\hline
 2370 & 2.723 & 18.28 & 11.21 & 3.086 & 18.06 & 11.27 & 0.151 & 24.56 & 11.36 \\\hline
 2380 & 2.658 & 18.28 & 11.26 & 3.013 & 18.07 & 11.32 & 0.1456 & 24.58 & 11.41 \\\hline
 2390 & 2.595 & 18.28 & 11.31 & 2.942 & 18.08 & 11.37 & 0.1404 & 24.6 & 11.46 \\\hline
 2400 & 2.534 & 18.28 & 11.37 & 2.873 & 18.08 & 11.43 & 0.1354 & 24.62 & 11.52 \\\hline
 2410 & 2.474 & 18.28 & 11.42 & 2.805 & 18.09 & 11.48 & 0.1306 & 24.65 & 11.58 \\\hline
 2420 & 2.415 & 18.28 & 11.47 & 2.74 & 18.1 & 11.54 & 0.126 & 24.67 & 11.63 \\\hline
 2430 & 2.358 & 18.28 & 11.53 & 2.675 & 18.1 & 11.59 & 0.1216 & 24.7 & 11.69 \\\hline
 2440 & 2.303 & 18.29 & 11.58 & 2.613 & 18.11 & 11.65 & 0.1173 & 24.72 & 11.74 \\\hline
 2450 & 2.248 & 18.29 & 11.64 & 2.552 & 18.12 & 11.7 & 0.1132 & 24.75 & 11.8 \\\hline
 2460 & 2.196 & 18.29 & 11.69 & 2.492 & 18.13 & 11.75 & 0.1092 & 24.77 & 11.85 \\\hline
 2470 & 2.144 & 18.29 & 11.75 & 2.434 & 18.13 & 11.81 & 0.1054 & 24.79 & 11.91 \\\hline
 2480 & 2.094 & 18.29 & 11.8 & 2.378 & 18.14 & 11.86 & 0.1017 & 24.82 & 11.97 \\\hline
 2490 & 2.045 & 18.3 & 11.86 & 2.323 & 18.15 & 11.92 & 0.0982 & 24.84 & 12.02 \\\hline
 2500 & 1.998 & 18.3 & 11.91 & 2.269 & 18.16 & 11.98 & 0.09479 & 24.87 & 12.08 \\\hline
 2510 & 1.951 & 18.3 & 11.97 & 2.216 & 18.17 & 12.03 & 0.09151 & 24.89 & 12.14 \\\hline
 2520 & 1.906 & 18.3 & 12.02 & 2.165 & 18.18 & 12.09 & 0.08834 & 24.92 & 12.19 \\\hline
 2530 & 1.862 & 18.31 & 12.08 & 2.115 & 18.18 & 12.14 & 0.08529 & 24.94 & 12.25 \\\hline
 2540 & 1.819 & 18.31 & 12.14 & 2.066 & 18.19 & 12.2 & 0.08235 & 24.97 & 12.31 \\\hline
 2550 & 1.777 & 18.31 & 12.19 & 2.019 & 18.2 & 12.26 & 0.07952 & 24.99 & 12.37 \\\hline
 2560 & 1.736 & 18.32 & 12.25 & 1.972 & 18.21 & 12.31 & 0.07679 & 25.02 & 12.42 \\\hline
 2570 & 1.696 & 18.32 & 12.31 & 1.927 & 18.21 & 12.37 & 0.07416 & 25.04 & 12.48 \\\hline
 2580 & 1.657 & 18.32 & 12.36 & 1.883 & 18.22 & 12.43 & 0.07163 & 25.06 & 12.54 \\\hline
 2590 & 1.619 & 18.32 & 12.42 & 1.84 & 18.23 & 12.48 & 0.06919 & 25.08 & 12.6 \\\hline
 2600 & 1.582 & 18.33 & 12.48 & 1.797 & 18.24 & 12.54 & 0.06683 & 25.11 & 12.66 \\\hline
 2610 & 1.545 & 18.33 & 12.53 & 1.756 & 18.25 & 12.6 & 0.06456 & 25.14 & 12.72 \\\hline
 2620 & 1.51 & 18.34 & 12.59 & 1.716 & 18.26 & 12.66 & 0.06237 & 25.17 & 12.78 \\\hline
 2630 & 1.476 & 18.34 & 12.65 & 1.677 & 18.27 & 12.72 & 0.06026 & 25.19 & 12.84 \\\hline
 2640 & 1.442 & 18.34 & 12.71 & 1.639 & 18.28 & 12.77 & 0.05822 & 25.22 & 12.9 \\\hline

\end{array}\nonumber
\end{equation}
\caption{\label{tab:xsterms6} NLO contributions to the production cross section for a scalar particle, as they are defined in eq.~\eqref{eq:master_bsm}. The theory uncertainty is computed as in eq.~\eqref{eq:NLO_err_incl_sc}. }
\end{table}

\begin{table}[!h]
\small
\begin{equation}
\begin{array}{|c|c|c|c||c|c|c||c|c|c|}
\hline
m_S \text{ [GeV]}
& \sigma^{NLO}_S[1,1] [\textrm{fb}]& \delta_{\text{th}}\,[\%]  &  \delta_{\alpha_S}^{\text{PDF}}\,[\%]
&  \sigma^{NLO}_S[1,0][\textrm{fb}]  &\delta_{\text{th}}\,[\%]   &  \delta_{\alpha_S}^{\text{PDF}}\,[\%]
&  \sigma^{NLO}_S[0,1][\textrm{fb}] & \delta_{\text{th}}\,[\%]    & \delta_{\alpha_S}^{\text{PDF}}\,[\%]
\\ \hline
 
 2650 & 1.409 & 18.35 & 12.77 & 1.602 & 18.28 & 12.83 & 0.05626 & 25.24 & 12.96 \\\hline
 2660 & 1.377 & 18.35 & 12.83 & 1.565 & 18.29 & 12.89 & 0.05436 & 25.27 & 13.02 \\\hline
 2670 & 1.346 & 18.36 & 12.88 & 1.53 & 18.3 & 12.95 & 0.05254 & 25.29 & 13.08 \\\hline
 2680 & 1.315 & 18.36 & 12.94 & 1.495 & 18.31 & 13.01 & 0.05077 & 25.31 & 13.14 \\\hline
 2690 & 1.286 & 18.36 & 13. & 1.461 & 18.32 & 13.07 & 0.04907 & 25.34 & 13.2 \\\hline
 2700 & 1.257 & 18.37 & 13.06 & 1.428 & 18.33 & 13.13 & 0.04743 & 25.37 & 13.26 \\\hline
 2710 & 1.228 & 18.38 & 13.12 & 1.396 & 18.34 & 13.19 & 0.04585 & 25.4 & 13.32 \\\hline
 2720 & 1.201 & 18.38 & 13.18 & 1.364 & 18.34 & 13.25 & 0.04432 & 25.42 & 13.38 \\\hline
 2730 & 1.174 & 18.38 & 13.24 & 1.333 & 18.35 & 13.31 & 0.04285 & 25.44 & 13.44 \\\hline
 2740 & 1.147 & 18.39 & 13.3 & 1.303 & 18.37 & 13.37 & 0.04143 & 25.47 & 13.5 \\\hline
 2750 & 1.121 & 18.4 & 13.36 & 1.274 & 18.37 & 13.43 & 0.04005 & 25.5 & 13.57 \\\hline
 2760 & 1.096 & 18.4 & 13.42 & 1.245 & 18.39 & 13.49 & 0.03873 & 25.53 & 13.63 \\\hline
 2770 & 1.072 & 18.41 & 13.48 & 1.217 & 18.39 & 13.55 & 0.03745 & 25.55 & 13.69 \\\hline
 2780 & 1.048 & 18.42 & 13.54 & 1.19 & 18.4 & 13.61 & 0.03621 & 25.58 & 13.75 \\\hline
 2790 & 1.024 & 18.42 & 13.61 & 1.163 & 18.41 & 13.68 & 0.03502 & 25.6 & 13.82 \\\hline
 2800 & 1.001 & 18.43 & 13.67 & 1.137 & 18.42 & 13.74 & 0.03387 & 25.63 & 13.88 \\\hline
 2810 & 0.9791 & 18.43 & 13.73 & 1.112 & 18.43 & 13.8 & 0.03276 & 25.65 & 13.94 \\\hline
 2820 & 0.9573 & 18.44 & 13.79 & 1.087 & 18.44 & 13.86 & 0.03169 & 25.69 & 14.01 \\\hline
 2830 & 0.936 & 18.45 & 13.85 & 1.062 & 18.45 & 13.92 & 0.03065 & 25.71 & 14.07 \\\hline
 2840 & 0.9152 & 18.45 & 13.92 & 1.039 & 18.46 & 13.99 & 0.02965 & 25.74 & 14.13 \\\hline
 2850 & 0.8949 & 18.46 & 13.98 & 1.016 & 18.47 & 14.05 & 0.02868 & 25.76 & 14.2 \\\hline
 2860 & 0.8751 & 18.46 & 14.04 & 0.9929 & 18.48 & 14.11 & 0.02775 & 25.79 & 14.26 \\\hline
 2870 & 0.8557 & 18.47 & 14.1 & 0.9708 & 18.49 & 14.17 & 0.02685 & 25.82 & 14.33 \\\hline
 2880 & 0.8368 & 18.48 & 14.17 & 0.9492 & 18.5 & 14.24 & 0.02598 & 25.84 & 14.39 \\\hline
 2890 & 0.8183 & 18.5 & 14.23 & 0.9281 & 18.51 & 14.3 & 0.02514 & 25.87 & 14.46 \\\hline
 2900 & 0.8003 & 18.51 & 14.29 & 0.9075 & 18.52 & 14.37 & 0.02432 & 25.9 & 14.52 \\\hline
 2910 & 0.7826 & 18.52 & 14.36 & 0.8874 & 18.53 & 14.43 & 0.02354 & 25.92 & 14.59 \\\hline
 2920 & 0.7654 & 18.53 & 14.42 & 0.8677 & 18.54 & 14.49 & 0.02278 & 25.95 & 14.65 \\\hline
 2930 & 0.7485 & 18.55 & 14.49 & 0.8485 & 18.55 & 14.56 & 0.02205 & 25.98 & 14.72 \\\hline
 2940 & 0.7321 & 18.56 & 14.55 & 0.8297 & 18.56 & 14.62 & 0.02134 & 26.01 & 14.79 \\\hline
 2950 & 0.716 & 18.57 & 14.62 & 0.8113 & 18.57 & 14.69 & 0.02065 & 26.03 & 14.85 \\\hline
 2960 & 0.7003 & 18.59 & 14.68 & 0.7934 & 18.58 & 14.75 & 0.01999 & 26.06 & 14.92 \\\hline
 2970 & 0.6849 & 18.6 & 14.75 & 0.7759 & 18.6 & 14.82 & 0.01935 & 26.09 & 14.99 \\\hline
 2980 & 0.6699 & 18.61 & 14.81 & 0.7588 & 18.6 & 14.89 & 0.01873 & 26.11 & 15.05 \\\hline
 2990 & 0.6552 & 18.63 & 14.88 & 0.742 & 18.62 & 14.95 & 0.01813 & 26.15 & 15.12 \\\hline
 3000 & 0.6409 & 18.64 & 14.95 & 0.7257 & 18.63 & 15.02 & 0.01756 & 26.18 & 15.19 \\\hline

 \end{array}\nonumber
\end{equation}
\caption{\label{tab:xsterms7a} NLO contributions to the production cross section for a scalar particle, as they are defined in eq.~\eqref{eq:master_bsm}. The theory uncertainty is computed as in eq.~\eqref{eq:NLO_err_incl_sc}. }
\end{table}

\begin{table}[!h]
\small
\begin{equation}
\begin{array}{|c|c|c|c||c|c|c||c|c|c|}
\hline
m_S \text{ [GeV]}
& \sigma^{NLO}_S[1,1] [\textrm{fb}] & \delta_{\text{th}}\,[\%]  &  \delta_{\alpha_S}^{\text{PDF}}\,[\%]
&  \sigma^{NLO}_S[1,0][\textrm{fb}]  &\delta_{\text{th}}\,[\%]   &  \delta_{\alpha_S}^{\text{PDF}}\,[\%]
&  \sigma^{NLO}_S[0,1][\textrm{fb}] & \delta_{\text{th}}\,[\%]    & \delta_{\alpha_S}^{\text{PDF}}\,[\%]
\\ \hline
 
 730 & 1.273 & 20.7 & 4.34 & 0.4417 & 18.35 & 4.38 & 0.5144 & 21.21 & 4.33 \\\hline
 731 & 1.264 & 20.7 & 4.34 & 0.4398 & 18.34 & 4.39 & 0.5105 & 21.21 & 4.33 \\\hline
 732 & 1.255 & 20.7 & 4.35 & 0.4379 & 18.34 & 4.39 & 0.5067 & 21.22 & 4.34 \\\hline
 733 & 1.246 & 20.7 & 4.35 & 0.4359 & 18.34 & 4.39 & 0.5029 & 21.22 & 4.34 \\\hline
 734 & 1.238 & 20.7 & 4.35 & 0.434 & 18.34 & 4.4 & 0.4991 & 21.22 & 4.34 \\\hline
 735 & 1.229 & 20.7 & 4.36 & 0.4321 & 18.33 & 4.4 & 0.4953 & 21.22 & 4.35 \\\hline
 736 & 1.221 & 20.7 & 4.36 & 0.4302 & 18.33 & 4.4 & 0.4916 & 21.23 & 4.35 \\\hline
 737 & 1.212 & 20.69 & 4.36 & 0.4283 & 18.33 & 4.41 & 0.4879 & 21.23 & 4.35 \\\hline
 738 & 1.204 & 20.69 & 4.37 & 0.4264 & 18.33 & 4.41 & 0.4843 & 21.23 & 4.36 \\\hline
 739 & 1.195 & 20.69 & 4.37 & 0.4246 & 18.33 & 4.41 & 0.4806 & 21.23 & 4.36 \\\hline
 740 & 1.187 & 20.69 & 4.37 & 0.4227 & 18.32 & 4.42 & 0.477 & 21.23 & 4.36 \\\hline
 741 & 1.179 & 20.69 & 4.38 & 0.4209 & 18.32 & 4.42 & 0.4735 & 21.24 & 4.37 \\\hline
 742 & 1.171 & 20.69 & 4.38 & 0.419 & 18.32 & 4.42 & 0.4699 & 21.24 & 4.37 \\\hline
 743 & 1.163 & 20.69 & 4.38 & 0.4172 & 18.32 & 4.43 & 0.4664 & 21.24 & 4.37 \\\hline
 744 & 1.155 & 20.69 & 4.39 & 0.4154 & 18.32 & 4.43 & 0.4629 & 21.24 & 4.38 \\\hline
 745 & 1.147 & 20.69 & 4.39 & 0.4136 & 18.31 & 4.43 & 0.4595 & 21.24 & 4.38 \\\hline
 746 & 1.139 & 20.69 & 4.4 & 0.4118 & 18.31 & 4.44 & 0.4561 & 21.24 & 4.39 \\\hline
 747 & 1.131 & 20.69 & 4.4 & 0.41 & 18.31 & 4.44 & 0.4527 & 21.25 & 4.39 \\\hline
 748 & 1.123 & 20.69 & 4.4 & 0.4082 & 18.31 & 4.44 & 0.4493 & 21.25 & 4.39 \\\hline
 749 & 1.115 & 20.68 & 4.4 & 0.4064 & 18.31 & 4.45 & 0.446 & 21.25 & 4.4 \\\hline
 750 & 1.108 & 20.68 & 4.41 & 0.4046 & 18.3 & 4.45 & 0.4427 & 21.25 & 4.4 \\\hline
 751 & 1.1 & 20.68 & 4.41 & 0.4029 & 18.3 & 4.45 & 0.4394 & 21.25 & 4.4 \\\hline
 752 & 1.093 & 20.68 & 4.42 & 0.4011 & 18.3 & 4.46 & 0.4361 & 21.26 & 4.41 \\\hline
 753 & 1.085 & 20.68 & 4.42 & 0.3994 & 18.3 & 4.46 & 0.4329 & 21.26 & 4.41 \\\hline
 754 & 1.078 & 20.68 & 4.42 & 0.3977 & 18.3 & 4.46 & 0.4297 & 21.26 & 4.41 \\\hline
 755 & 1.07 & 20.68 & 4.43 & 0.3959 & 18.29 & 4.47 & 0.4265 & 21.26 & 4.42 \\\hline
 756 & 1.063 & 20.68 & 4.43 & 0.3942 & 18.29 & 4.47 & 0.4233 & 21.26 & 4.42 \\\hline
 757 & 1.056 & 20.67 & 4.43 & 0.3925 & 18.29 & 4.47 & 0.4202 & 21.26 & 4.42 \\\hline
 758 & 1.049 & 20.67 & 4.44 & 0.3908 & 18.29 & 4.48 & 0.4171 & 21.26 & 4.43 \\\hline
 759 & 1.042 & 20.67 & 4.44 & 0.3892 & 18.28 & 4.48 & 0.414 & 21.27 & 4.43 \\\hline
 760 & 1.035 & 20.67 & 4.44 & 0.3875 & 18.28 & 4.49 & 0.411 & 21.27 & 4.43 \\\hline
 761 & 1.028 & 20.67 & 4.45 & 0.3858 & 18.28 & 4.49 & 0.408 & 21.27 & 4.44 \\\hline
 762 & 1.021 & 20.67 & 4.45 & 0.3842 & 18.28 & 4.49 & 0.405 & 21.27 & 4.44 \\\hline
 763 & 1.014 & 20.67 & 4.45 & 0.3825 & 18.28 & 4.5 & 0.402 & 21.28 & 4.44 \\\hline
 764 & 1.007 & 20.67 & 4.46 & 0.3809 & 18.27 & 4.5 & 0.399 & 21.28 & 4.45 \\\hline
 765 & 1. & 20.67 & 4.46 & 0.3792 & 18.27 & 4.5 & 0.3961 & 21.28 & 4.45 \\\hline
 766 & 0.9934 & 20.67 & 4.46 & 0.3776 & 18.27 & 4.51 & 0.3932 & 21.28 & 4.45 \\\hline
 767 & 0.9867 & 20.67 & 4.47 & 0.376 & 18.27 & 4.51 & 0.3903 & 21.28 & 4.46 \\\hline
 768 & 0.9801 & 20.66 & 4.47 & 0.3744 & 18.27 & 4.51 & 0.3874 & 21.29 & 4.46 \\\hline
 769 & 0.9735 & 20.66 & 4.48 & 0.3728 & 18.27 & 4.52 & 0.3846 & 21.29 & 4.46 \\\hline
 770 & 0.967 & 20.66 & 4.48 & 0.3712 & 18.26 & 4.52 & 0.3818 & 21.29 & 4.47 \\\hline

 \end{array}\nonumber
\end{equation}
\caption{\label{tab:xsterms7} NLO contributions to the production cross section for a scalar particle with a mass around $750$~GeV, as they are defined in eq.~\eqref{eq:master_bsm}. The theory uncertainty is computed as in eq.~\eqref{eq:NLO_err_incl_sc}. }
\end{table}

\end{document}